\documentclass [11pt]{article}
\usepackage{amsfonts}
\usepackage{graphicx}
\usepackage{amsmath,amssymb}
\usepackage{latexsym}
\usepackage{mathrsfs}
\usepackage{bbold}
\usepackage{color}
\usepackage{slashed}
\definecolor{red}{rgb}{1,0,0}
\usepackage{cancel}
\usepackage{comment}
\usepackage[nosort]{cite}
\usepackage{xcolor}
\usepackage{multirow}
\definecolor{red}{rgb}{1,0,0}

\makeatletter
\def\section{\@startsection {section}{1}{\z@}{-3.5ex plus -1ex minus
 -.2ex}{2.3ex plus .2ex}{\large\bf}}
\def\subsection{\@startsection{subsection}{2}{\z@}{-3.25ex plus -1ex
minus -.2ex}{1.5ex plus .2ex}{\normalsize\bf}}
\makeatother
\makeatletter

\@addtoreset{equation}{section}

\makeatother
\def\bel{\begin{equation}\begin{aligned}}
\def\eel{\end{aligned}\end{equation}}

\def\bea{\begin{eqnarray}} \def\eea{\end{eqnarray}}
\def\be{\begin{equation}} \def\ee{\end{equation}} \def\nn{\nonumber}

\newcommand{\promille}{%
  \relax\ifmmode\promillezeichen
        \else\leavevmode\(\mathsurround=0pt\promillezeichen\)\fi}
\newcommand{\promillezeichen}{%
  \kern-.05em%
  \raise.5ex\hbox{\the\scriptfont0 0}%
  \kern-.15em/\kern-.15em%
  \lower.25ex\hbox{\the\scriptfont0 00}}

\usepackage[colorlinks,linkcolor=black,citecolor=blue,urlcolor=blue,linktocpage]{hyperref}

\begin{document}

\thispagestyle{empty}

\begin{center}

\vspace*{-.6cm}

\hfill SISSA 21/2015/FISI \\

\begin{center}

\vspace*{1.1cm}

{\Large\bf Deconstructing Conformal Blocks in 4D CFT}

\end{center}

\vspace{0.8cm}

{\bf Alejandro Castedo Echeverri$^{a}$, Emtinan Elkhidir$^{a}$,\\[3mm]}
{\bf  Denis Karateev$^{a}$, Marco Serone$^{a,b}$}

\vspace{1.cm}

${}^a\!\!$
{\em SISSA and INFN, Via Bonomea 265, I-34136 Trieste, Italy}

\vspace{.1cm}

${}^b\!\!$
{\em ICTP, Strada Costiera 11, I-34151 Trieste, Italy}

\end{center}

\vspace{1cm}

\centerline{\bf Abstract}
\vspace{2 mm}
\begin{quote}

We show how  conformal partial waves (or conformal blocks) of spinor/tensor correlators can be related to each other by means of  differential operators  in four dimensional conformal field theories. We explicitly construct such differential operators for all possible conformal partial waves associated to four-point functions of arbitrary
traceless symmetric operators. Our method allows any conformal partial wave to  be extracted from a few ``seed" correlators, simplifying dramatically the computation needed to
bootstrap tensor correlators.

\end{quote}
 
%\vfill

\newpage

\tableofcontents

\section{Introduction}

There has been a revival of interest  in recent years in  four dimensional (4D) Conformal Field Theories (CFTs), after the seminal paper \cite{Rattazzi:2008pe} resurrected the old idea of the bootstrap program \cite{Ferrara:1973yt,Polyakov:1974gs}.  A 4D CFT is determined in terms of its spectrum of primary operators and the coefficients entering three-point functions among such primaries. Once this set of CFT data is given, any correlator is in principle calculable.  Starting from this observation, ref.\cite{Rattazzi:2008pe} has shown how imposing crossing symmetry
in four point functions can lead to non-trivial sets of constraints on the CFT data.
These are based on first principles and apply to any CFT, with or without a Lagrangian description.
Although any correlator can in principle be ``bootstrapped", in practice one has to be able to sum, for each primary operator exchanged in the correlator in some kinematical channel, the contribution of its infinite series of descendants. Such contribution is often called a conformal block. In fact, the crucial technical ingredient in ref.\cite{Rattazzi:2008pe} was the work of refs.\cite{Dolan:2000ut,Dolan:2003hv}, where  such conformal blocks have been explicitly computed for scalar four-point functions. Quite remarkably, the authors of refs.\cite{Dolan:2000ut,Dolan:2003hv} were able  to pack the contributions of traceless symmetric operators  of any spin  into a very simple formula. 

Since ref.\cite{Rattazzi:2008pe}, there have been many developments, both analytical \cite{Heemskerk:2009pn,Fitzpatrick:2011ia,Costa:2011mg,Costa:2011dw,Maldacena:2011jn,SimmonsDuffin:2012uy,Pappadopulo:2012jk,Fitzpatrick:2012yx,Komargodski:2012ek,Hogervorst:2013sma,Hogervorst:2013kva,Behan:2014dxa,Alday:2014tsa,Costa:2014rya,Vos:2014pqa,Elkhidir:2014woa,Goldberger:2014hca,Kaviraj:2015cxa,Alday:2015eya,Kaviraj:2015xsa}  and numerical \cite{Rychkov:2009ij,Caracciolo:2009bx,Poland:2010wg,Rattazzi:2010gj,Rattazzi:2010yc,Vichi:2011ux,Poland:2011ey,Liendo:2012hy,Beem:2013qxa,Gliozzi:2013ysa,Alday:2013opa,Berkooz:2014yda,Alday:2014qfa,Caracciolo:2014cxa,Beem:2014zpa,Bobev:2015jxa}
in the 4D bootstrap. All numerical studies are still based on identical scalar correlators,  unless  supersymmetry or global symmetries are present.\footnote{The techniques to bootstrap correlators with non identical fields were developed in refs.\cite{Kos:2014bka,Simmons-Duffin:2015qma}.
They have been used so far in 3D only, although they clearly apply in any number of space-time dimensions.}
There is an obvious reason for this limitation.
Determining the conformal blocks relevant for four-point functions involving tensor primary operators is significantly more complicated. First of all, contrary to their scalar counterpart,
tensor four-point correlators are determined in terms of several functions, one for each independent allowed tensor structure. Their number $N_4$ grows very rapidly with the spin of the external operators. The whole contribution of primary operators in any given channel is no longer parametrized by a single conformal block as in the scalar case, but in general by $N_4\times N_4$ conformal blocks, $N_4$ for each independent tensor structure. For each exchanged primary operator,  it is  convenient not to talk of individual conformal blocks but of Conformal Partial Waves (CPW), namely the entire contribution given by several conformal blocks, one for each tensor structure. Second, the exchanged operator is no longer necessarily traceless symmetric, but can be in an arbitrary representation of the 4D Lorentz group, depending  on the external operators and on the channel considered.

CPW can be determined in terms of the product of two three-point functions, each involving two external operators and the exchanged one.
If it is possible to relate a three-point function to another simpler one, a relation between CPW associated to different four-point functions can be obtained. Using this simple observation,  building on previous work \cite{Costa:2011mg}, in ref.\cite{Costa:2011dw} the CPW associated to a correlator of traceless symmetric operators (in arbitrary space-time dimensions), which exchange a traceless symmetric operator, have been related to the scalar conformal block of refs.\cite{Dolan:2000ut,Dolan:2003hv}. Despite this significant progress, bootstrapping tensor four-point functions in 4D requires the knowledge of the CPW associated to the exchange of non-traceless symmetric operators. Even for traceless symmetric exchange, the methods of refs.\cite{Costa:2011mg,Costa:2011dw} do not allow to study  correlators with external non-traceless symmetric fields (although generalizations that might do that  have been proposed, see ref.\cite{Costa:2014rya}).

The aim of this paper is to make a step forward and generalize the relation between  CPW  found for traceless symmetric operators in ref.\cite{Costa:2011dw}
to arbitrary CPW in 4D CFTs. We will perform this task by using the 6D embedding formalism in terms of twistors. 
Our starting point is the recent general classification of 3-point functions found in ref.\cite{Elkhidir:2014woa}. We will see how  three-point functions of spinors/tensors can be related
to three-point functions of lower spin fields by means of differential operators.  We explicitly construct a basis of differential operators that allows one to express
any three-point function of two traceless symmetric and an arbitrary bosonic operator ${\cal O}^{l,\bar l}$ with $l\neq \bar l$, in terms of  ``seed" three-point functions, that admit a unique tensor structure.
This would allow to express all the CPW entering a four-point function of traceless symmetric correlators in terms of a few CPW seeds.
We do not attempt to compute such seeds explicitly, although it might be done by developing the methods of refs.\cite{Dolan:2000ut,Dolan:2003hv}. 

The structure of the paper is as follows. In section 2 we will briefly review the 6D embedding formalism in twistor space in index-free notation and the results of ref.\cite{Elkhidir:2014woa} on the three-point function classification. In section 3 we recall how a relation between three-point functions leads to a relation between CPW.
We introduce our differential operators in section 4. We construct an explicit basis of differential operators in section 5 for external symmetric traceless operators.
In subsection 5.1  we reproduce (and somewhat improve) the results of ref.\cite{Costa:2011dw} in our formalism where the exchanged operator is traceless symmetric and then pass to the more involved case of mixed tensor exchange in subsection 5.2.
In section 6 we discuss the basis of the tensor structures of four-point functions and propose a set of  seed CPW needed to get CPW associated with the exchange of a bosonic operator ${\cal O}^{l,\bar l}$.  
A couple of examples are proposed in section 7. 
In subsection 7.1 we consider a four fermion correlator and in subsection 7.2 we schematically deconstruct spin one and spin two correlators, and show how to impose their conservation.  
We conclude in section 8, where we discuss in particular the computations yet to be done to bootstrap tensor correlators in 4D CFTs.
A (non-exhaustive) list of relations between $SU(2,2)$ invariants entering four-point functions is listed in appendix A.

\section{Three-Point Function Classification}
\label{sec:class}

General three-point functions in 4D CFTs involving bosonic or fermionic operators in irreducible representations of the Lorentz group have recently been classified and computed in ref.\cite{Elkhidir:2014woa} (see refs.\cite{Osborn:1993cr,Erdmenger:1996yc} for important early works on tensor correlators and refs.\cite{Weinberg:2010fx,Costa:2011mg,Costa:2011dw,Stanev:2012nq,Zhiboedov:2012bm,Dymarsky:2013wla,Costa:2014rya,Li:2014gpa,Korchemsky:2015ssa} for other recent studies) using the 6D embedding formalism \cite{Dirac:1936fq,Mack:1969rr,Ferrara2,Dobrev:1977qv} formulated in terms of twistors in an index-free notation \cite{SimmonsDuffin:2012uy} (see e.g. refs.\cite{Siegel:1992ic,Siegel:2012di,Goldberger:2011yp,Goldberger:2012xb,Fitzpatrick:2014oza,Khandker:2014mpa} for applications mostly in the context of supersymmetric CFTs).
We will here briefly review the main results of ref.\cite{Elkhidir:2014woa}. 

A 4D primary operator ${\cal O}_{\alpha_1\ldots \alpha_l}^{\dot{\beta}_1\ldots \dot{\beta}_{\bar l}}$ with scaling dimension $\Delta$ in the $(l,\bar l)$ representation of the Lorentz group can be embedded in a 6D multi-twistor field $O^{a_1\ldots a_l}_{b_1\ldots b_{\bar l}}$, homogeneous of degree $n=\Delta+ (l+\bar l)/2$, as follows:
\be
{\cal O}_{\alpha_1\ldots \alpha_l}^{\dot{\beta}_1\ldots \dot{\beta}_{\bar l}}(x) =   (X^+)^{\Delta-(l+\bar l)/2}  \mathbf{X}_{\alpha_1 a_1}\ldots  \mathbf{X}_{\alpha_l a_l}  \overline{\mathbf{X}}^{\dot\beta_1 b_1}
\ldots  \overline{\mathbf{X}}^{\dot\beta_{\bar l} b_{\bar l}} O^{a_1\ldots a_l}_{b_1\ldots b_{\bar l}}(X)\,.
\label{fFrelation}
\ee
In eq.(\ref{fFrelation}), 6D and 4D coordinates are denoted as $X^M$ and $x^\mu$, where $x^\mu = X^\mu/X^+$, ${\mathbf{X}}$ and $\overline{\mathbf{X}}$ are 6D twistor space-coordinates defined as
\be
\mathbf{X}_{ab} \equiv  X_M \Sigma^M_{ab} = - \mathbf{X}_{ba} \,, \ \ \ \overline{\mathbf{X}}^{ab}  \equiv X_M \overline\Sigma^{Mab}= - \overline{\mathbf{X}}^{ba}\,,
\label{TwistorCoord}
\ee
in terms of the 6D chiral Gamma matrices $\Sigma^M$ and $\overline{\Sigma}^M$ (see Appendix A of  ref.\cite{Elkhidir:2014woa} for further details).
One has $\mathbf{X} \overline{\mathbf{X}} =\overline{\mathbf{X}} \mathbf{X}  =X_MX^M=X^2$, which vanishes on the null 6D cone.

It is very useful to use an index-free notation by defining
\be
O(X,S,\bar S) \equiv  O^{a_1\ldots a_l}_{b_1\ldots b_{\bar l}}(X)\ S_{a_1} \ldots S_{a_l} \bar S^{b_1} \ldots \bar S^{b_{\bar l}} \,.
\label{Findexfree}
\ee
A 4D field ${\cal O}$ is actually uplifted to an equivalence class of 6D fields $O$. Any two fields $O$ and $\hat O= O+ \overline{\mathbf{X}} V$
or $\hat O= O+ {\mathbf{X}} \overline{W}$, for some multi twistors $V$ and $\overline{W}$, are equivalent uplifts of ${\cal O}$.

Given a 6D multi-twistor field $O$, the corresponding 4D field ${\cal O}$ is obtained by taking
\be
 {\cal O}_{\alpha_1\ldots \alpha_l}^{\dot{\beta}_1\ldots \dot{\beta}_{\bar l}}(x) = \frac{ (X^+)^{\Delta-\frac{l+\bar l}{2}} }{l!\bar l!} \Big( {\mathbf{X}}\frac{\partial}{\partial S}\Big)_{\alpha_1} \ldots  \Big( {\mathbf{X}}\frac{\partial}{\partial S}\Big)_{\alpha_l} \Big( {\overline{\mathbf{X}}}\frac{\partial}{\partial \bar S}\Big)^{\dot\beta_1}\ldots \Big( {\overline{\mathbf{X}}}\frac{\partial}{\partial \bar S}\Big)^{\dot\beta_{\bar l}}
O\Big(X,S,\bar S\Big) \,.
  \label{f4dExp}
\ee
The 4D three-point functions are conveniently encoded in their scalar 6D counterpart $\langle O_1 O_2 O_3 \rangle$ which must be a sum of $SU(2,2)$ invariant quantities
constructed out of the  $X_i$, $S_i$ and $\bar S_i$, with the correct homogeneity properties under rescaling.
Notice that quantities proportional to $\bar S_i \mathbf{X}_i$, $\overline{\mathbf{X}}_i S_i$ or $\bar S_i S_i$ ($i=1,2,3$) are projected to zero in 4D.
The non-trivial $SU(2,2)$ possible invariants are ($i\neq j\neq k$, indices not summed) \cite{SimmonsDuffin:2012uy}:
\begin{align}
\label{eq:invar1}
I_{ij}   & \equiv \bar S_i S_j  \,, \\
\label{eq:invar2}
K_{i,jk}            &\equiv N_{i,jk} S_j \overline{\mathbf{X}}_i S_k \,, \\
\label{eq:invar3}
\overline{K}_{i,jk} &\equiv N_{i,jk} \bar S_j \mathbf{X}_i \bar S_k \,,\\
\label{eq:invar4}
J_{i,jk}            &\equiv  N_{jk} \bar S_i  \mathbf{X}_j \overline{\mathbf{X}}_k S_i \,,
\end{align}
where
\be
N_{jk}   \equiv \frac{1}{X_{jk}}\,, \ \ \
N_{i,jk}  \equiv \sqrt{\frac{X_{jk}}{X_{ij}X_{ik}}}\,.
\label{eq:Ninva}
\ee
Two-point functions are easily determined. One has
\be
\langle O_1(X_1,S_1,\bar S_1) O_2(X_2,S_2,\bar S_2) \rangle = X_{12}^{-\tau_1} I_{21}^{l_1} I_{12}^{\bar l_1} \delta_{l_1,\bar l_2} \delta_{l_2,\bar l_1} \delta_{\Delta_1, \Delta_2} \,,
\label{2ptFun}
\ee
where $X_{ij}\equiv X_i\cdot X_j$ and  $\tau_i\equiv \Delta_i+(l_i+\bar l_i)/2$. As can be seen from eq.(\ref{2ptFun}), any operator $O^{l,\bar l}$ has a non-vanishing two-point function with a conjugate operator $O^{\bar l,l}$ only. 

The main result of ref.\cite{Elkhidir:2014woa} can be recast in the following way.
The most general three-point function $\langle O_1 O_2 O_3 \rangle$ can be written as\footnote{The points $X_1$, $X_2$ and $X_3$ are assumed to be distinct.}
 \be\label{eq:ff3pf}
\langle O_1 O_2 O_3 \rangle= \sum_{s=1}^{N_3}  \lambda_s  \langle O_1 O_2 O_3 \rangle_s \,,
\ee
where
\be
  \langle O_1 O_2 O_3 \rangle_s = \mathcal{K}_3 \Big(\prod_{i\neq j=1}^3 I_{ij}^{m_{ij}} \Big) C_{1,23}^{n_1}C_{2,31}^{n_2}C_{3,12}^{n_3} \,.
\label{eq:ff3pfV2}
\ee
In eq.(\ref{eq:ff3pfV2}), $\mathcal{K}_3$ is a kinematic factor that depends on the scaling dimension and spin of the external fields,
\be\label{eq:kinematicfactor1}
\mathcal{K}_3=\frac{1}{X_{12}^{a_{12}} X_{13}^{a_{13}} X_{23}^{a_{23}}},
\ee
with $a_{ij} =(\tau_i+\tau_j-\tau_k)/2$,  $i\neq j\neq k$.
The index $s$ runs over all the independent tensor structures parametrized by the integers $m_{ij}$ and $n_i$, each multiplied by a constant OPE coefficient $\lambda_s$.
The invariants $C_{i,jk}$ equal to one of the three-index invariants (\ref{eq:invar2})-(\ref{eq:invar4}), depending on the value of
\be\label{eq:dl}
\Delta l\equiv l_1+l_2+l_3-(\bar l_1+\bar l_2+\bar l_3)\,,
\ee
of the external fields.
Three-point functions are non-vanishing only when $\Delta l$ is an even integer \cite{Mack:1976pa,Elkhidir:2014woa}. We have
\begin{itemize}
\item{$\Delta l=0$: $C_{i,jk}=J_{i,jk}$.} 
\item{$\Delta l>0$: $C_{i,jk}=J_{i,jk}, K_{i,jk}$.}
\item{$\Delta l<0$: $C_{i,jk}=J_{i,jk}, \overline K_{i,jk}$.}
\end{itemize}
A redundance is present for $\Delta l=0$. It can be fixed by demanding, for instance,  that one of the three integers $n_i$ in eq.(\ref{eq:ff3pfV2}) vanishes.
The total number of $K_{i,jk}$'s ($\overline K_{i,jk}$'s) present in the correlator for $\Delta l>0$ ($\Delta l<0$) equal $\Delta l/2$ ($-\Delta l/2$).
The number of tensor structures is given by all the possible allowed choices of nonnegative integers $m_{ij}$ and $n_i$ in eq.(\ref{eq:ff3pf}) subject to the above constraints and the ones
coming from matching the correct powers of $S_i$ and $\bar S_i$ for each field. The latter requirement gives in total six constraints.

Conserved 4D operators are encoded in multitwistors $O$ that satisfy the
current conservation condition
\be
D\cdot O(X,S,\bar S) = 0 \,, \ \ \ \ \
D = \Big(X_M \Sigma^{MN} \frac{\partial}{\partial X^N}\Big)_{a}^{\;b}  \frac{\partial}{\partial S_a} \frac{\partial}{\partial \bar S^b}\,.
\label{ConservedD}
\ee
When eq.(\ref{ConservedD}) is imposed on eq.(\ref{eq:ff3pf}), we generally get a set of linear relations between the OPE coefficients $\lambda_s$'s, which restrict the possible allowed tensor structures in the three point function. Under a 4D parity transformation, the invariants (\ref{eq:invar1})-(\ref{eq:invar4}) transform as follows:
\be\begin{split}
I_{ij} \stackrel{{\cal P}}{\longrightarrow}  \;\; & - I_{ji} \,,\\
K_{i,jk} \stackrel{{\cal P}}{\longrightarrow}  \;\;  & + \overline{K}_{i,jk} \,, \\
\overline{K}_{i,jk} \stackrel{{\cal P}}{\longrightarrow}  \;\;  & + K_{i,jk} \,, \\
 J_{i,jk} \stackrel{{\cal P}}{\longrightarrow}  \;\;  & + J_{i,jk} \,.
 \label{6Dparity}
\end{split}
\ee

\section{Relation between CPW}
\label{sec:cpw}

A CFT is defined in terms of the spectrum of primary operators, their scaling dimensions $\Delta_i$ and $SL(2,C)$ representations $(l_i,\bar l_i)$,  and OPE coefficients, namely the coefficients entering the three-point functions among such primaries. Once this set of CFT data is given, any correlator is in principle calculable. Let us consider for instance the 4-point function of four primary tensor operators:
\be
\langle {\cal O}^{I_1}_1(x_1) {\cal O}^{I_2}_2(x_2) {\cal O}^{I_3}_3(x_3) {\cal O}^{I_4}_4(x_4)\rangle = \mathcal{K}_4 \sum_{n=1}^{N_4} g_n(u,v) {\cal T}_n^{I_1I_2I_3I_4}(x_i) \,.
\label{Gen4pt}
\ee
In eq.(\ref{Gen4pt}) we have schematically denoted by $I_{i}$ the Lorentz indices of the operators ${\cal O}_i(x_i)$,  $x_{ij}^2=(x_i-x_j)_\mu(x_i-x_j)^\mu$,
\be\label{eq:4kinematic}
\mathcal{K}_4 = \bigg(\frac{x_{24}^2}{x_{14}^2}\bigg)^{\frac{\tau_1-\tau_2}2}
\bigg(\frac{x_{14}^2}{x_{13}^2}\bigg)^{\frac{\tau_3-\tau_4}{2}}
(x_{12}^2)^{-\frac{\tau_1+\tau_2}2} (x_{34}^2)^{-\frac{\tau_3+\tau_4}2} 
\ee
is a kinematical factor, $u$ and $v$ are the usual conformally invariant cross ratios
\be
u=\frac{x_{12}^2x_{34}^2}{x_{13}^2x_{24}^2}\,, \ \ \ v=\frac{x_{14}^2 x_{23}^2}{x_{13}^2x_{24}^2}\,,
\label{uv4d}
\ee
${\cal T}_n^{I_1I_2I_3I_4}(x_i)$ are tensor structures and $\tau_i$ are defined below eq.(\ref{2ptFun}). These are functions of the $x_i$'s and can be kinematically determined.
Their total number $N_4$ depends on the Lorentz properties of the external primaries. For correlators involving scalars only, one has  $N_4=1$, but in general
$N_4>1$ and rapidly grows with the spin of the external fields. For instance, for four traceless symmetric operators with identical spin $l$, one has $N_4(l)\sim l^7$ for large $l$ \cite{Elkhidir:2014woa}.
All the non-trivial dynamical information of the 4-point function is encoded in the $N_4$ functions $g_n(u,v)$.
In any given channel, by using the OPE we can write the 4-point function (\ref{Gen4pt}) in terms of the operators exchanged in that channel.
In the s-channel (12-34), for instance,  we have
\be
\langle {\cal O}^{I_1}_1(x_1) {\cal O}^{I_2}_2(x_2) {\cal O}^{I_3}_3(x_3) {\cal O}^{I_4}_4(x_4)\rangle =  \sum_r\sum_{p=1}^{N_{3r}^{12}} \sum_{q=1}^{N_{3\bar r}^{34}}\sum_{{\cal O}_r} \lambda_{{\cal O}_1{\cal O}_2{\cal O}_r}^p \lambda_{\bar {\cal O}_{\bar r}{\cal O}_3{\cal O}_4}^q W_{{\cal O}_1{\cal O}_2{\cal O}_3{\cal O}_4,{\cal O}_r}^{(p,q)I_1I_2I_3I_4}(x_i)\,,
\label{sch4pt}
\ee
where %$r$ runs over the allowed classes of  representations of the exchanged operator ${\cal O}_r$, 
$p$ and $q$ run over the possible independent tensor structures associated to the three point functions $\langle {\cal O}_1{\cal O}_2{\cal O}_r\rangle$ and $\langle {\bar {\cal O}}_{\bar r}{\cal O}_3{\cal O}_4\rangle$, whose total number is $N_{3r}^{12}$ and $N_{3\bar r}^{34}$ respectively,\footnote{Strictly speaking these numbers depend also on ${\cal O}_r$, particularly on its spin. When the latter is large enough, however, $N_{3r}^{12}$ and $N_{3\bar r}^{34}$ are only functions of the external operators.}  the $\lambda$'s being their corresponding structure constants, and  $r$ and ${\cal O}_r$ runs over the  number of primary operators that can be exchanged in the correlator. We divide the (infinite) sum over the exchanged operators in a {\it finite} sum over the different classes of representations that can appear, e.g.  $(l,l)$, $(l+2,l)$, etc., while the sum over ${\cal O}_r$
includes the sum over the scaling dimension and spin $l$ of the operator exchanged within the class $r$. For example, four-scalar correlators can only exchange traceless symmetric operators and hence the
sum over $r$ is trivial. Finally, in eq.(\ref{sch4pt}) $W_{{\cal O}_1{\cal O}_2{\cal O}_3{\cal O}_4}^{(p,q)I_1I_2I_3I_4}(u,v)$ are the so-called CPW associated to the four-point function. They depend on the external as well as the exchanged operator scaling dimension and spin, dependence we omitted in order not to clutter further the notation.\footnote{For further simplicity, in what follows we will often omit the subscript indicating the external operators associated to the CPW.}  By comparing eqs.(\ref{Gen4pt}) and (\ref{sch4pt}) one  can infer that the number of allowed tensor structures in three and four-point functions is related:\footnote{We do not have a formal proof of eq.(\ref{N4N3}), although the agreement found in ref.\cite{Elkhidir:2014woa} using eq.(\ref{N4N3}) in different channels is a strong indication that it should be correct.} 
\be
N_4=\sum_r N_{3r}^{12} N_{3\bar r}^{34}\,.
\label{N4N3}
\ee
There are several CPW for each exchanged primary operator ${\cal O}_r$, depending on the number of allowed 3-point function structures. They encode the contribution of all the descendant operators associated to the primary ${\cal O}_r$.
Contrary to the functions $g_n(u,v)$ in eq.(\ref{Gen4pt}), the CPW do not carry dynamical information, being determined by conformal symmetry alone.
They admit a parametrization like the 4-point function itself,
\be
W_{{\cal O}_1{\cal O}_2{\cal O}_3{\cal O}_4,{\cal O}_r}^{(p,q)I_1I_2I_3I_4}(x_i) = \mathcal{K}_4 \sum_{n=1}^{N_4} {\cal G}_{{\cal O}_r,n}^{(p,q)}(u,v)  {\cal T}_n^{I_1I_2I_3I_4}(x_i) \,,
\label{WGen}
\ee
where
${\cal G}^{(p,q)}_{{\cal O}_r,n}(u,v)$ are conformal blocks depending on $u$ and $v$ and on the dimensions and spins of the external and exchanged operators.
Once the CPW are  determined,  by comparing eqs.(\ref{Gen4pt}) and (\ref{sch4pt}) we can express $g_n(u,v)$ in terms of the OPE coefficients of the exchanged operators.
This procedure can be done in other channels as well, $(13-24)$ and $(14-23)$. Imposing crossing symmetry by requiring  the equality of different channels is the essence of the bootstrap approach.

The computation of CPW of tensor correlators is possible, but technically is not easy. In particular it is desirable to have a relation between different CPW, so that it is enough to compute a small subset of them, 
which determines all the others. In order to understand how this reduction process works,  it is very useful to embed the CPW in the 6D embedding space with an index-free notation. We use here the formalism in terms of twistors as reviewed in section \ref{sec:class}.
It is useful to consider the parametrization of CPW in the shadow formalism \cite{Ferrara:1972xe,Ferrara:1972uq,Ferrara:1972ay, Ferrara:1973vz}.  It has been shown in ref.\cite{SimmonsDuffin:2012uy} that a generic CPW can be written in 6D as
\be
W_{O_1 O_2 O_3 O_4,O_r}^{(p,q)}(X_i) \propto \! \int d^4Xd^4Y  \langle  O_1(X_1) O_2(X_2) O_r(X,S,\bar S)\rangle_p G
 \langle  \bar O_{\bar r}(Y,T,\bar T) O_3(X_3) O_4(X_4)\rangle_q \,.
 \label{shadow}
\ee
In eq.(\ref{shadow}), $O_i(X_i)=O_i(X_i,S_i,\bar S_i)$ are the index-free 6D fields associated to the 4D fields ${\cal O}_i(x_i)$,  $O_r(X,S,\bar S)$ and $\bar O_{\bar r}(Y,T,\bar T)$  are the exchanged operator and its conjugate, $G$ is a sort of ``propagator", function of $X,Y$ and of the twistor derivatives $\partial/\partial S$, $\partial/\partial T$, $\partial/\partial \bar S$ and $\partial/\partial \bar T$, and the subscripts $p$ and $q$ label the three-point function tensor structures. Finally, in order to remove unwanted contributions, the transformation $X_{12}\rightarrow  e^{4\pi i} X_{12}$ should be performed and the integral should be projected to the suitable eigenvector under the above monodromy. We do not provide additional details, which can be found in ref.\cite{SimmonsDuffin:2012uy}, since they are irrelevant for our considerations.  Suppose one is able to find a relation between three-point functions of this form:
\be
 \langle  O_1(X_1) O_2(X_2) O_r(X,S,\bar S)\rangle_p = D_{pp^\prime}(X_{12},S_{1,2},\bar S_{1,2}) \langle O_1^\prime(X_1) O_2^\prime(X_2) O_r(X,S,\bar S)\rangle_{p^\prime}\,,
 \label{OOOpp}
 \ee
where $D_{pp^\prime}$ is some operator that depends on $X_{12},S_{1,2},\bar S_{1,2}$ and their derivatives, but is crucially  independent of $X$, $S$, and $\bar S$, and $O_i^\prime(X_i)$ are some other, possibly  simpler, tensor operators. As long as the operator  $D_{pp^\prime}(X_{12},S_{1,2},\bar S_{1,2})$ does not change the monodromy properties of the integral,  one can use eq.(\ref{OOOpp}) in both three-point functions entering eq.(\ref{shadow}) and move the operator $D_{pp^\prime}$ outside the integral. In this way we get, with obvious notation,
\be
W_{O_1 O_2 O_3 O_4,O_r}^{(p,q)}(X_i) =   D_{pp^\prime}^{12}  D_{qq^\prime}^{34} W_{O_1^\prime O_2^\prime O_3^\prime O_4^\prime,O_r}^{(p^\prime,q^\prime)}(X_i)  \,.
 \label{shadow2}
\ee
Using the embedding formalism in vector notation, ref.\cite{Costa:2011dw} has shown how to reduce,  in any space-time dimension, CPW
associated to a correlator of traceless symmetric operators which exchange a traceless symmetric operator to the known CPW of scalar correlators \cite{Dolan:2000ut,Dolan:2003hv}.

Focusing on 4D CFTs and using the embedding formalism in twistor space,  we will see how the reduction of CPW  can be generalized for arbitrary external and exchanged operators.

\section{Differential Representation of Three-Point Functions}
\label{sec:OB}

We look for an explicit expression of the operator $D_{pp^\prime}$ defined in eq.(\ref{OOOpp}) as a linear combination of products of simpler operators. 
They must raise (or more generically change) the degree in $S_{1,2}$ and have to respect the gauge redundancy we have in the choice of $O$. As we recalled in subsection \ref{sec:class},
multitwistors $O$ and $\hat O$ of the form
\be
\hat O = O +  (\bar S X) G + (\overline X S)  G'\,, \ \ \ \ \ \hat O = O + (X^2) G \,,
\label{F1}
\ee
where $G$ and $G'$ are some other multi-twistors fields, are equivalent uplifts of the same 4D tensor field. Eq.(\ref{OOOpp}) is gauge invariant with respect to the equivalence classes  (\ref{F1})
only if we demand
\begin{equation}
D_{pp^\prime}(\mathbf{\overline X}_i\mathbf{X}_i,\mathbf{\overline X}_i S_i, \overline S_i\mathbf{X}_i, X_i^2,\overline S_i S_i )\propto (\mathbf{\overline X}_i\mathbf{X}_i,\mathbf{\overline X}_i S_i, \overline S_i\mathbf{X}_i, X_i^2, \overline S_i S_i)\,, \ \ i=1,2\,.
\label{Consist}
\end{equation}

It is useful to classify the building block operators according  to their value of $\Delta l$, as defined in eq.(\ref{eq:dl}).

At zero order in derivatives, we have three possible operators, with $\Delta l = 0$:
\be
\sqrt{X_{12}}, I_{12}\,, I_{21}\,.
\label{I12I21}
\ee
At first order in derivatives  (in $X$ and $S$), four operators are possible with $\Delta l=  0$:
\begin{equation}\begin{aligned}
D_1&\equiv\frac{1}{2}\overline S_1 \Sigma^M \overline\Sigma^N S_1\Big(X_{2M}\frac{\partial}{\partial X^N_1}-X_{2N}\frac{\partial}{\partial X^M_1}\Big)\,, \\
D_2&\equiv\frac{1}{2}\overline S_2 \Sigma^M \overline\Sigma^N S_2\Big(X_{1M}\frac{\partial}{\partial X^N_2}-X_{1N}\frac{\partial}{\partial X^M_2}\Big)\,, \\
\widetilde D_1&\equiv\overline S_1 \mathbf{X}_2 \overline\Sigma^N S_1\frac{\partial}{\partial X^N_2}+2I_{12}\,S_{1a}\frac{\partial}{\partial S_{2a}}-2I_{21}\,\overline S^a_1\frac{\partial}{\partial\overline S^a_2}\,,\\
\widetilde D_2&\equiv\overline S_2 \mathbf{X}_1 \overline\Sigma^N S_2\frac{\partial}{\partial X^N_1}+2I_{21}\,S_{2a}\frac{\partial}{\partial S_{1a}}-2I_{12}\,\overline S^a_2\frac{\partial}{\partial\overline S^a_1}\,.\\
\end{aligned}
\label{DDtilde}
\end{equation}
The extra two terms in the last two lines of eq.(\ref{DDtilde}) are needed to satisfy the condition (\ref{Consist}).
The $SU(2,2)$ symmetry forbids any operator at first order in derivatives with $\Delta l=  \pm 1$.

When $\Delta l=  2$, we have the two operators
\begin{equation}\begin{aligned}
d_{1} \equiv S_2 \overline X_{1} \frac{\partial}{\partial\overline S_1}\,,   \ \ \ \ \ \ 
d_{2}  \equiv S_1 \overline X_{2} \frac{\partial}{\partial\overline S_2}\,,   
\end{aligned}
\label{D12}
\end{equation}
and their conjugates with  $\Delta l= - 2$:
\begin{equation}\begin{aligned}
\overline d_{1}\equiv \overline S_2 X_{1} \frac{\partial}{\partial S_1}\,, \ \ \ \ \ \ \
\overline d_{2}\equiv  \overline S_1 X_{2} \frac{\partial}{\partial S_2}\,.
\end{aligned}
\label{Dbar12}
\end{equation}
The operator $\sqrt{X_{12}}$ just decreases the dimensions at both points 1 and 2 by one half.
The operator $I_{12}$  increases by one the spin $\bar l_{1}$ and by one $l_{2}$.
The operator $D_{1}$  increases by one the spin $l_{1}$ and by one $\bar l_{1}$, increases by one the dimension at point 1 and decreases by one the dimension at point 2.
The operator $\widetilde D_1$ increases by one the spin $l_{1}$ and by one the spin $\bar l_{1}$ and it does not change the dimension of both points 1 and 2.
The operator $d_1$ increases by one the spin $l_{2}$ and decreases by one $\bar l_{1}$,  decreases by one the dimension at point 1 and does not change the dimension at point 2.
The action of the remaining operators is trivially obtained by $1\leftrightarrow 2$ exchange or by conjugation.

Two more operators with $\Delta l=2$ are possible:
\begin{equation}\begin{aligned}
\widetilde d_{1}& \equiv   X_{12}  S_1\overline\Sigma^M S_2\frac{\partial}{\partial X^N_1}-I_{12}S_{1a}\mathbf{\overline X}_2^{ab}\frac{\partial}{\partial\overline S^b_1}\,, \\
\widetilde d_{2}& \equiv   X_{12} S_2\overline\Sigma^M S_1\frac{\partial}{\partial X^N_2}- I_{21}S_{2a}\mathbf{\overline X}_1^{ab}\frac{\partial}{\partial\overline S^b_2}\,,
\end{aligned}
\label{Dtilde12}
\end{equation}
together with their conjugates with $\Delta l=-2$. We will shortly see that the operators (\ref{Dtilde12}) are redundant and can be neglected.

The above operators satisfy the commutation relations
\begin{equation}\begin{aligned}
& [D_i,\widetilde D_j ] = [d_i,d_j] = [\bar d_i,\bar d_j] =  [d_i,\widetilde d_j] = [\bar d_i,\overline{\widetilde d}_j] =[\widetilde d_i,\widetilde d_j] = [\overline{\widetilde d}_i,\overline{\widetilde d}_j] = 0\,,\ \ \ \    i,j=1,2  \,, \\
& [D_1, D_2 ] = 4 I_{12} I_{21} \Big( -X_1^M \frac{\partial}{\partial X_1^M}+X_2^M \frac{\partial}{\partial X_2^M} \Big)\,, \\
& [\widetilde D_1, \widetilde D_2 ] = 4 I_{12} I_{21} \Big( X_1^M \frac{\partial}{\partial X_1^M}-X_2^M \frac{\partial}{\partial X_2^M}+S_1 \frac{\partial}{\partial S_1}+\bar S_1 \frac{\partial}{\partial \bar S_1}-S_2 \frac{\partial}{\partial S_2}-\bar S_2 \frac{\partial}{\partial \bar S_2} \Big)\,, \\
& [\widetilde d_1,\overline{\widetilde d}_2] = 2 X_{12} I_{12} I_{21} \Big(- X_1^M \frac{\partial}{\partial X_1^M}+X_2^M \frac{\partial}{\partial X_2^M}-\bar S_1 \frac{\partial}{\partial \bar S_1}+S_2 \frac{\partial}{\partial S_2} \Big)\,, \\ 
& [d_i, \bar d_j ] = 2X_{12} \Big(S_j \frac{\partial}{\partial S_j}-\bar S_i \frac{\partial}{\partial \bar S_i}\Big) (1-\delta_{i,j}) \,, \ \ \ \   i,j=1,2\,,  \\
&[ d_i, D_j] = -2 \delta_{i,j} \widetilde d_i \,,\ \  \ \  i,j=1,2 \,, \\
&[ d_1, \widetilde D_1] = 2 \widetilde d_2 \,,\hspace{1.9cm} [ d_2, \widetilde D_1] = 0\,,  \\
& [\widetilde d_1,D_1]=0\,, \hspace{2.3cm}  [\widetilde d_2,D_1]=-2 I_{12} I_{21} d_2\,, \\
& [\widetilde d_1,\widetilde D_1]=2 I_{12} I_{21} d_2 \,, \hspace{.95cm}  [\widetilde d_2,\widetilde D_1]=0 \,, \\
& [d_1,\overline{\widetilde d}_1] =-X_{12} \widetilde D_2 \,, \hspace{1.27cm}  [d_1,\overline{\widetilde d}_2] =  X_{12} D_2 \,.
\end{aligned}
\label{commutators}
\end{equation}
Some other commutators are trivially obtained by exchanging 1 and 2 and by the parity transformation (\ref{parity}).  The operators $\sqrt{X}_{12}$, $I_{12}$ and $I_{21}$ commute
with all the differential operators. Acting on the whole correlator, we have
\begin{equation}
S_i \frac{\partial}{\partial S_i} \rightarrow l_i \,, \ \ \  \bar S_i \frac{\partial}{\partial \bar S_i} \rightarrow \bar l_i  \,, \ \ \ \
 X_i^M \frac{\partial}{\partial X_i^M} \rightarrow - \tau_i \,,
\label{Xssaction}
\end{equation}
and hence the above differential operators, together with $X_{12}$ and $I_{12} I_{21}$, form a closed algebra when acting on three-point correlators.
Useful information on conformal blocks can already be obtained by considering the rather trivial operator $\sqrt{X_{12}}$. For any three point function tensor structure, we have
\be
\langle O_1 O_2 O_3 \rangle_s = (\sqrt{X_{12}})^{a} \langle O_1^{\frac a2} O_2^{\frac a2} O_3 \rangle_s\,,
\label{X12n}
\ee
where $a$ is an integer (in order not to induce a monodromy for $X_{12}\rightarrow  e^{4\pi i} X_{12}$) and the superscript indicates a shift in dimension. If $\Delta({\cal O})=\Delta_{\cal O}$, then $\Delta({\cal O}^a)=\Delta_{\cal O}+a$. Using eqs.(\ref{X12n}) and (\ref{shadow2}), we get
for any 4D CPW and pair of integers $a$ and $b$:
\be
W_{{\cal O}_1 {\cal O}_2 {\cal O}_3 {\cal O}_4,{\cal O}_r}^{(p,q)} =    x_{12}^{a}   x_{34}^{b} W_{{\cal O}_1^{\frac a2} {\cal O}_2^{\frac a2}  {\cal O}_3^{\frac b2}  {\cal O}_4^{\frac b2} ,{\cal O}_r}^{(p,q)}  \,.
 \label{shadow2M}
\ee
In terms of the conformal blocks defined in eq.(\ref{WGen}) one has
\be
{\cal G}_{{\cal O}_r,n}^{(p,q)}(u,v)={\cal G}_{{\cal O}_r,n}^{(p,q)\frac a2,\frac a2,\frac b2,\frac b2}(u,v) \,,
\label{Gshifts}
\ee
where the superscripts indicate the shifts in dimension in the four external operators. Equation (\ref{Gshifts}) significantly constrains the dependence of ${\cal G}_{{\cal O}_r,n}^{(p,q)}$
on the external operator dimensions $\Delta_i$. The conformal blocks can be periodic functions of $\Delta_1$, $\Delta_2$ and $\Delta_3$, $\Delta_4$, but
can arbitrarily depend on $\Delta_1-\Delta_2$, $\Delta_3-\Delta_4$. This is in agreement with the known form of scalar conformal blocks. Since in this paper we are mostly concerned in deconstructing tensor structures, we will neglect in the following the operator $\sqrt{X_{12}}$.

The set of differential operators is redundant, namely there is generally more than 1 combination of products of operators that lead from one three-point function structure to another one. 
In particular, without any loss of generality we can forget about the operators (\ref{Dtilde12}), since their action is equivalent to commutators of $d_i$ and $D_j$.
On the other hand, it is not difficult to argue that the above operators do not allow to connect any three-point function structure to any other one.
For instance, it is straightforward to verify that there is no way  to connect a three-point correlator with one $(l,\bar l)$ field to another correlator
with a ($l\pm1,\bar l \mp 1$) field, with the other fields left unchanged. This is not an academic observation because, as we will see, connections of this kind will turn out to be useful
in order to simplify the structure of the CPW seeds.
The problem is solved by adding to the above list of operators the following second-order operator with $\Delta l=0$:
\be
\nabla_{12} \equiv  \frac{(\overline{\mathbf X}_1{\mathbf X}_2)^a_b}{X_{12}}\frac{\partial^2}{\partial\overline{S}_1^a\partial S_{2,b}}
\label{nablaS}
\ee
and its conjugate $\nabla_{21}$. 
The above operators transform as follows under 4D parity:
\begin{equation}
D_i \stackrel{{\cal P}}{\longrightarrow} D_i \,, \ \ \ \ \   \widetilde D_i \stackrel{{\cal P}}{\longrightarrow}  \widetilde  D_i \,, \ \ \  d_i  \stackrel{{\cal P}}{\longleftrightarrow}- \overline d_i \,, \ \ \ \ 
 \widetilde d_i  \stackrel{{\cal P}}{\longleftrightarrow}   \widetilde{\overline d}_i \,,  \ \ (i=1,2)\,, \ \ \ \nabla_{12} \stackrel{{\cal P}}{\longleftrightarrow} -\nabla_{21}\,.
\label{parity}
\end{equation}
It is clear that all the operators above are invariant under the monodromy $X_{12}\rightarrow  e^{4\pi i} X_{12}$. 
The addition of $\nabla_{12}$ and $\nabla_{21}$ makes the operator basis even more redundant.
It is clear that the paths connecting two different three-point correlators that make use of the least number of these operators are preferred, in particular those that also avoid (if possible)
the action of  the second order operators $\nabla_{12}$ and $\nabla_{21}$.
We will not attempt here to explicitly construct a minimal differential basis connecting two arbitrary three-point correlators. Such an analysis is in general complicated and perhaps not really necessary, since in most applications we are interested in CPW involving external fields with spin up to two.
Given their particular relevance, we will instead focus in the next section on three-point correlators of two traceless symmetric operators with an arbitrary field $O^{(l,\bar l)}$.

\section{Differential Basis for Traceless Symmetric Operators}
\label{sec:DBTSO}

In this section we show how three-point correlators of two traceless symmetric operators with an arbitrary field $O^{(l_3,\bar l_3)}$ can be reduced to seed correlators,
with one tensor structure only. We first consider the case $l_3=\bar l_3$, and then go on with $l_3 \neq \bar l_3$.

\subsection{Traceless Symmetric Exchanged Operators}
\label{subsec:DBTSO}

The reduction of traceless symmetric correlators to lower spin traceless symmetric correlators has been successfully addressed in ref.\cite{Costa:2011dw}.
In this subsection  we essentially reformulate the results of ref.\cite{Costa:2011dw} in our formalism. This will turn out to be crucial to address the more complicated case of mixed symmetry operator exchange. Whenever possible, we will use a notation as close as possible to that of ref.\cite{Costa:2011dw}, in order to make any comparison more transparent to the reader.

Three-point correlators of traceless symmetric operators can be expressed only in terms of the $SU(2,2)$ invariants $I_{ij}$ and $J_{i,jk}$ defined in eqs.(\ref{eq:invar1})-(\ref{eq:invar4}),
since $\Delta l$ defined in eq.(\ref{eq:dl}) vanishes. It is useful to consider separately parity even and parity odd tensor structures.
Given the action of parity, eq.(\ref{6Dparity}), the most general parity even tensor structure is given by products of the following invariants:
\begin{equation}
\label{RedBasisST}
(I_{21}I_{13}I_{32}-I_{12}I_{31}I_{23}), (I_{12}I_{21}), (I_{13}I_{31}), (I_{23}I_{32}), J_{1,23},J_{2,31},J_{3,12}\,.
\end{equation}
These structures are not all independent, because of the identity
\begin{equation}
J_{1,23}J_{2,31}J_{3,12}=8(I_{12}I_{31}I_{23}-I_{21}I_{13}I_{32})-4(I_{23}I_{32}J_{1,23}+I_{13}I_{31}J_{2,31}+I_{12}I_{21}J_{3,12}) \,.
\label{J3Rel}
\end{equation}
In ref.\cite{Elkhidir:2014woa},  eq.(\ref{J3Rel}) has been used to define an independent basis where no tensor structure contains the three $SU(2,2)$ invariants  $J_{1,23}$, $J_{2,31}$ and $J_{3,12}$ at the same time. A more symmetric and convenient basis is obtained by using eq.(\ref{J3Rel}) to get rid of the first factor in eq.(\ref{RedBasisST}).
We define the most general parity even tensor structure of traceless symmetric tensor correlator as
\begin{equation}
\left[\begin{array}{ccc}\Delta_1&\Delta_2&\Delta_3\\l_1&l_2&l_3\\m_{23}&m_{13}&m_{12}\end{array}\right] \equiv  \mathcal{K}_3
(I_{12}I_{21})^{m_{12}}(I_{13}I_{31})^{m_{13}}(I_{23}I_{32})^{m_{23}}J_{1,23}^{j_1}J_{2,31}^{j_2}J_{3,12}^{j_3} \,,
\label{ParityevenInd}
\end{equation}
where $l_i$ and $\Delta_i$ are the spins and scaling dimensions of the fields, the kinematical factor $\mathcal{K}_3$ is defined in eq.(\ref{eq:kinematicfactor1}) and
\begin{equation}
\begin{array}{l}j_1=l_1-m_{12}-m_{13}\geq 0 \,, \\ j_2=l_2-m_{12}-m_{23} \geq 0 \,, \\ j_3=l_3-m_{13}-m_{23} \geq 0 \,. \end{array}
\label{j123}
\end{equation}
Notice the similarity of eq.(\ref{ParityevenInd}) with eq.(3.15) of ref.\cite{Costa:2011dw}, with $(I_{ij}I_{ji})\rightarrow H_{ij}$ and $J_{i,jk}\rightarrow V_{i,jk}$.
The structures (\ref{ParityevenInd}) can be related to a seed scalar-scalar-tensor correlator. Schematically
\begin{equation}
\left[\begin{array}{ccc}\Delta_1&\Delta_2&\Delta_3\\l_1&l_2&l_3\\m_{23}&m_{13}&m_{12}\end{array}\right] = {\cal  D}
\left[\begin{array}{ccc}\Delta_1^\prime&\Delta_2^\prime&\Delta_3\\0 & 0&l_3\\ 0 & 0 & 0 \end{array}\right]   \,,
\label{OBeven}
\end{equation}
where ${\cal  D}$ is a sum of products of the operators introduced in section \ref{sec:OB}. Since symmetric traceless correlators have $\Delta l=0$, it is natural to expect that only the
operators with $\Delta l=0$ defined in eqs.(\ref{I12I21}) and (\ref{DDtilde}) will enter in ${\cal D}$. Starting from the seed, we now show how one can iteratively construct all tensor structures by means of recursion relations. The analysis will be very similar to the one presented in ref.\cite{Costa:2011dw} in vector notation.
We first construct tensor structures with $m_{13}=m_{32}=0$ for any $l_1$ and $l_2$ by iteratively using the relation (analogue of  eq.(3.27) in ref.\cite{Costa:2011dw}, with $D_1\rightarrow D_{12}$
and $\widetilde D_1\rightarrow D_{11}$)
\begin{equation}\begin{aligned}
 & D_1\left[\begin{array}{ccc}\Delta_1&\Delta_2+1&\Delta_3\\l_1-1&l_2&l_3\\0&0&m_{12}\end{array}\right]+\tilde D_1\left[\begin{array}{ccc}\Delta_1+1&\Delta_2&\Delta_3\\l_1-1&l_2&l_3\\0&0&m_{12}\end{array}\right] = \\
& (2+2m_{12}-l_1-l_2-\Delta_3)\left[\begin{array}{ccc}\Delta_1&\Delta_2&\Delta_3\\l_1&l_2&l_3\\0&0&m_{12}\end{array}\right]
-8(l_2-m_{12})\left[\begin{array}{ccc}\Delta_1&\Delta_2&\Delta_3\\l_1&l_2&l_3\\0&0&m_{12}+1\end{array}\right] \,.
\end{aligned}
\label{RR1}
\end{equation}
The analogous equation with $D_2$ and $\widetilde D_2$ is obtained from eq.(\ref{RR1}) by exchanging $1\leftrightarrow 2$ and changing sign of the coefficients in the right hand side of the equation. The sign change arises from the fact that $J_{1,23}\rightarrow -J_{2,31}$,  $J_{2,31}\rightarrow -J_{1,23}$ and $J_{3,12}\rightarrow -J_{3,12}$ under $1\leftrightarrow 2$. Hence structures that differ by one spin get a sign change. This observation applies also to eq.(\ref{RR2}) below.
Structures with $m_{12}>0$ are deduced using (analogue of eq(3.28) in ref.\cite{Costa:2011dw})
\begin{equation}
\left[\begin{array}{ccc}\Delta_1&\Delta_2&\Delta_3\\l_1&l_2&l_3\\m_{23}&m_{13}&m_{12}\end{array}\right]= (I_{12}I_{21})\left[\begin{array}{ccc} \Delta_1+1& \Delta_2+1&\Delta_3\\l_1-1&l_2-1&l_3\\m_{23}&m_{13}&m_{12}-1\end{array}\right] \,.
\end{equation}
Structures with non-vanishing $m_{13}$ ($m_{23}$) are obtained by acting with the operator $D_{1}$ ($D_2$):
\begin{equation}
\begin{array}{l}4(l_3-m_{13}-m_{23}) \left[\begin{array}{ccc}\Delta_1&\Delta_2&\Delta_3\\l_1&l_2&l_3\\m_{23}&m_{13}+1&m_{12}\end{array}\right]= D_1\left[\begin{array}{ccc}\Delta_1&\Delta_2+1&\Delta_3\\l_1-1&l_2&l_3\\m_{23}&m_{13}&m_{12}\end{array}\right]\\+4(l_2-m_{12}-m_{23})\left[\begin{array}{ccc}\Delta_1&\Delta_2&\Delta_3\\l_1&l_2&l_3\\m_{23}&m_{13}&m_{12}+1\end{array}\right]-\\
\frac{1}{2}(2+2m_{12}-2m_{13}+\Delta_2-\Delta_1-\Delta_3-l_1-l_2+l_3)\left[\begin{array}{ccc}\Delta_1&\Delta_2&\Delta_3\\l_1&l_2&l_3\\m_{23}&m_{13}&m_{12}\end{array}\right] \,, \end{array}
\label{RR2}
\end{equation}
and is the analogue of eq (3.29) in ref.\cite{Costa:2011dw}. In this way all parity even tensor structures can be constructed starting from the seed correlator.

Let us now turn to parity odd structures. The most general parity odd structure is given by
\begin{equation}
\left[\begin{array}{ccc}\Delta_1&\Delta_2&\Delta_3\\l_1&l_2&l_3\\m_{23}&m_{13}&m_{12}\end{array}\right]_{odd} \equiv
(I_{12}I_{23} I_{31}+I_{21}I_{32} I_{13})
\left[\begin{array}{ccc}\Delta_1+1&\Delta_2+1&\Delta_3+1\\l_1-1&l_2-1&l_3-1\\m_{23}&m_{13}&m_{12}\end{array}\right]   \,.
\label{ParityoddInd}
\end{equation}
Since the parity odd combination $(I_{12}I_{23} I_{31}+I_{21}I_{32} I_{13})$ commutes with $D_{1,2}$ and $\widetilde D_{1,2}$, the recursion relations found for parity even structures
straightforwardly apply to the parity odd ones.
One could define a ``parity odd seed" 
\begin{equation}
16 l_3 (\Delta_3-1) \left[\begin{array}{ccc}\Delta_1&\Delta_2&\Delta_3\\1&1&l_3\\0&0& 0\end{array}\right]_{odd} =(d_2 \bar d_1-\bar d_2 d_1)D_1D_2
\left[\begin{array}{ccc}\Delta_1+2&\Delta_2+2&\Delta_3\\0 & 0&l_3\\ 0 & 0 & 0 \end{array}\right]   
\label{parityoddseed}
\end{equation}
and from here construct all the parity odd structures. Notice that the parity odd seed cannot be obtained by applying only combinations of $D_{1,2}$, $\widetilde D_{1,2}$ and $(I_{12}I_{21})$, because
these operators are all invariant under parity, see eq.(\ref{parity}). This explains the appearance of the operators $d_i$ and $\bar d_i$ in eq.(\ref{parityoddseed}).
The counting of parity even and odd structures manifestly agrees with that performed in ref.\cite{Costa:2011mg}.

Once proved that all tensor structures can be reached by acting with operators on the seed correlator, one might define a differential basis which is essentially identical to that defined in eq.(3.31) of ref. \cite{Costa:2011dw}:
\be
{\small{ \left\{\begin{array}{ccc}\Delta_1&\Delta_2&\Delta_3\\l_1&l_2&l_3\\m_{23}&m_{13}&m_{12}\end{array}\right\}_{\!\! 0} }= (I_{12} I_{21})^{m_{12}} D_1^{m_{13}} D_2^{m_{23}} \widetilde D_1^{j_1} \widetilde D_2^{j_2}
{\small\left[\begin{array}{ccc}\Delta_1^\prime&\Delta_2^\prime&\Delta_3\\0 & 0&l_3\\ 0 & 0 & 0 \end{array}\right]} }\,,
\label{DBeven}
\ee
where $\Delta_1^\prime=\Delta_1+l_1+m_{23}-m_{13}$, $\Delta_2^\prime=\Delta_2+l_2+m_{13}-m_{23}$.
The recursion relations found above have shown that the differential basis (\ref{DBeven}) is  complete: all parity even tensor structures can be written as linear combinations of eq.(\ref{DBeven}).  The dimensionality of the differential basis matches  the one of the ordinary basis for any spin $l_1$, $l_2$ and $l_3$.
Since both bases are complete, the  transformation matrix relating them is ensured to have maximal rank. Its determinant, however, is a function of the scaling dimensions $\Delta_i$ and the spins $l_i$ of the fields and one should check that it does not vanish for some specific values of $\Delta_i$ and $l_i$. We have explicitly checked up to $l_1=l_2=2$ that for $l_3\geq l_1+l_2$ the rank of the transformation matrix depends only on $\Delta_3$ and $l_3$ and never vanishes, for any value of $\Delta_3$ allowed by the unitarity bound \cite{Mack:1975je}.
On the other hand, a problem can arise when $l_3<l_1+l_2$, because in this case a dependence on the values of $\Delta_1$ and $\Delta_2$ arises and the determinant vanishes for specific values (depending on the $l_i$'s) of $\Delta_1-\Delta_2$ and $\Delta_3$, even when they are within the unitarity bounds.\footnote{A similar problem seems also to occur for the basis (3.31) of ref. \cite{Costa:2011dw} in vector notation.}  
This issue is easily solved by replacing $\widetilde D_{1,2}\rightarrow (\widetilde D_{1,2}+D_{1,2})$ in eq.(\ref{DBeven}), as suggested by the recursion relation (\ref{RR1}), and by defining an improved differential basis
\begin{equation}\begin{aligned}
& {\small{\left\{\begin{array}{ccc}\Delta_1&\Delta_2&\Delta_3\\l_1&l_2&l_3\\m_{23}&m_{13}&m_{12}\end{array}\right\} }}= (I_{12} I_{21})^{m_{12}} D_1^{m_{13}} D_2^{m_{23}} \!\!\sum_{n_1=0}^{j_1}\!\!\Big(
\begin{array}{c} j_1 \\ n_1\end{array}\Big) D_1^{n_1} \widetilde D_1^{j_1-n_1}
\!\!\sum_{n_2=0}^{j_2}\!\!\Big(
\begin{array}{c} j_2 \\ n_2\end{array}\Big) D_2^{n_2} \widetilde D_2^{j_2-n_2}
{\small{\left[\begin{array}{ccc}\Delta_1^{\prime}&\Delta_2^{\prime} &\Delta_3\\0 & 0&l_3\\ 0 & 0 & 0 \end{array}\right] } }
\label{DBevenNewST}
\end{aligned}
\end{equation}
where $\Delta_1^{\prime} =  \Delta_1 + l_1+m_{23}-m_{13}+n_2-n_1$, $\Delta_2^{\prime} =  \Delta_2 + l_2+m_{13}-m_{23}+n_1-n_2$. 
A similar basis for parity odd structures is given by
\be
{\small{ \left\{\begin{array}{ccc}\Delta_1&\Delta_2&\Delta_3\\l_1&l_2&l_3\\m_{23}&m_{13}&m_{12}\end{array}\right\}_{odd} }} \!\!\! =(d_2 \bar d_1-\bar d_2 d_1)D_1 D_2
 {\small{\left\{\begin{array}{ccc}\Delta_1+2 &\Delta_2+2&\Delta_3\\ l_1-1 & l_2-1 &l_3\\ m_{23} & m_{13} & m_{12} \end{array}\right\} }} \,.
 \label{DBodd}
\ee
In practical computations it is more convenient to use the differential basis rather than the recursion relations and, if necessary,  use the transformation matrix to rotate the results back to the ordinary basis.
We have explicitly constructed the improved differential basis (\ref{DBevenNewST}) and (\ref{DBodd})  up to  $l_1=l_2=2$. The rank of the transformation matrix depends on $\Delta_3$ and $l_3$ for any value of $l_3$, and never vanishes, for any value of $\Delta_3$ allowed by the unitary bound.\footnote{The transformation matrix is actually not of maximal rank when $l_3=0$ and $\Delta_3=1$.  However, this case is quite trivial. The exchanged scalar is free and hence the CFT is the direct sum of at least two CFTs, the interacting one and the free theory associated to this scalar. So, either the two external $l_1$ and $l_2$ tensors  are part of the free CFT, in which case the whole correlator is  determined, or the OPE coefficients entering the correlation function must vanish.\label{foot:1}}

\subsection{Mixed Symmetry Exchanged Operators}
\label{subsec:DBAO}

In this subsection we consider correlators with two traceless symmetric and one mixed symmetry operator $O^{(l_3,\bar l_3)}$,
with $l_3-\bar l_3=2\delta$, with $\delta$ an integer. A correlator of this form has $\Delta l= 2 \delta$ and
according to the analysis of section \ref{sec:class}, any of its tensor structures can be expressed in a form containing an overall number $\delta$ of $K_{i,jk}$'s if $\delta>0$, or
$\overline K_{i,jk}$'s if $\delta<0$. We consider in the following $\delta>0$,  the case $\delta<0$ being easily deduced from $\delta>0$ by means of
a parity transformation. The analysis will proceed along the same lines of subsection \ref{subsec:DBTSO}.
We first show a convenient parametrization for the tensor structures of the correlator, then we prove by deriving recursion relations how all tensor structures can be reached starting from a single seed,
to be determined, and finally present a differential basis.

We first consider the situation where $l_3\geq l_1+l_2-\delta$ and then the slightly more involved case with unconstrained $l_3$.

\subsubsection{Recursion Relations for $l_3\geq l_1+l_2-\delta$}
\label{subsubsec:largel3}

It is convenient to look for a parametrization of the tensor structures which is as close as possible to the one (\ref{ParityevenInd}) valid for $\delta=0$.
When $l_3\geq l_1+l_2-\delta$, any tensor structure of the correlator contains  enough $J_{3,12}$'s invariants to
remove all possible $K_{3,12}$'s invariants using the identity
\begin{equation}\label{JK3}
J_{3,12} K_{3,12} =2 I_{31} K_{1,23} - 2 I_{32} K_{2,31}\,.
\end{equation}
There are four possible combinations in which the remaining $K_{1,23}$ and $K_{2,31}$ invariants can enter in the correlator: 
$K_{1,23} I_{23}$, $K_{1,23} I_{21} I_{13}$ and  $K_{2,31} I_{13}$, $K_{2,31} I_{12} I_{23}$.
These structures are not all independent. In addition to eq.(\ref{JK3}), using the two identities
\begin{equation}\begin{aligned}\label{JK12}
2 I_{12} K_{2,31} &= J_{1,23} K_{1,23} + 2 I_{13} K_{3,12}  \,, \\
2 I_{21} K_{1,23} &= - J_{2,31} K_{2,31}  + 2 I_{23} K_{3,12} \,,
\end{aligned}\end{equation}
we can remove half of them and keep only, say, $K_{1,23} I_{23}$ and $K_{2,31} I_{13}$.
The most general tensor structure can be written as
\begin{equation}{\small{
\left[\begin{array}{ccc}\Delta_1&\Delta_2&\Delta_3\\l_1&l_2&l_3\\m_{23}&m_{13}&m_{12}\end{array}\right] _p \equiv  \Big(\frac{K_{1,23} I_{23}}{X_{23}}\Big)^{\delta-p}
 \Big(\frac{K_{2,31} I_{13}}{X_{13}}\Big)^{p}
 \left[\begin{array}{ccc}\Delta_1&\Delta_2&\Delta_3\\l_1-p&l_2-\delta+p &l_3\\m_{23}&m_{13}&\widetilde m_{12}\end{array}\right] \,, \ \ \ p=0,\ldots,\delta \,,}}
\label{deltaGenExpST}
\end{equation}
expressed in terms of the parity even structures (\ref{ParityevenInd}) of traceless symmetric correlators, where
\begin{equation}\begin{aligned}\label{j123AO}
j_1 & = l_1-p - \widetilde m_{12}- m_{13} \geq 0 \,, \\ j_2 & = l_2-\delta+p -\widetilde m_{12}-m_{23}\geq 0 \,, \\
j_3 & =l_3 -m_{13}-m_{23} \geq 0
\end{aligned} \hspace{2cm} 
\begin{aligned}
\widetilde m_{12}  = \left\{\begin{array}{l}m_{12} \,\,\,\,\;\;\; if \;\;\;\;\;p=0 \,\,\,or \,\,\, p= \delta \\ 0 \,\,\,\,\,\,\,\,\, \,\;\;otherwise \end{array}\right.    \,.
\end{aligned}
\end{equation}
The condition in $m_{12}$ derives from the fact that, using eqs.(\ref{JK12}), one can set $m_{12}$ to zero in the tensor structures with $p\neq 0,\delta$, see below.
Attention should be paid to the subscript $p$. Structures with no subscript refer to purely traceless symmetric correlators, while those with the subscript $p$ refer to three-point functions with two traceless symmetric
and one mixed symmetry field. All tensor structures are classified in terms of $\delta+1$ classes,
parametrized by the index $p$ in eq.(\ref{deltaGenExpST}). The parity odd structures of traceless symmetric correlators do not enter, since they can be reduced in the form (\ref{deltaGenExpST}) by means of the identities (\ref{JK12}). The class $p$ exists only when $l_1\geq p$ and $l_2\geq \delta-p$. If $l_1+l_2<\delta$, the entire correlator vanishes.

Contrary to the symmetric traceless exchange, there is no obvious choice of seed that stands out.
The allowed correlator with the lowest possible spins in each class, $l_1=p$, $l_2=\delta-p$, $m_{ij}=0$, can all be seen as possible seeds
with a unique tensor structure.  Let us see how all the structures (\ref{deltaGenExpST}) can be iteratively constructed using the operators defined in section \ref{sec:OB} in terms of the $\delta+1$ seeds.  It is convenient to first construct a redundant basis where $m_{12}\neq 0$ for any $p$ and then impose the relation that leads to the independent basis (\ref{deltaGenExpST}).
The procedure is similar to that followed for the traceless symmetric exchange.  We first construct all the tensor structures with $m_{13}=m_{32}=0$ for any spin $l_1$ and $l_2$, and any class $p$, using the following relations:
 {\small{\bea
&& D_1 \left[\begin{array}{ccc}\Delta_1&\Delta_2+1&\Delta_3\\l_1-1&l_2&l_3\\ 0& 0&m_{12}\end{array}\right] _p+
\widetilde D_1 \left[\begin{array}{ccc}\Delta_1+1&\Delta_2&\Delta_3\\l_1-1&l_2&l_3\\0& 0&m_{12}\end{array}\right] _p = (\delta-p) \left[\begin{array}{ccc}\Delta_1&\Delta_2&\Delta_3\\l_1&l_2&l_3\\ 0& 0&m_{12}\end{array}\right] _{p+1}   \\
&& \!\!\!\!  -8(l_2-\delta+p-m_{12})   \left[\begin{array}{ccc}\Delta_1&\Delta_2&\Delta_3\\l_1&l_2&l_3\\ 0& 0&m_{12}+1\end{array}\right] _{p}+(2m_{12}-l_1-l_2-\Delta_3+2+\delta-p)  \left[\begin{array}{ccc}\Delta_1&\Delta_2&\Delta_3\\l_1&l_2&l_3\\ 0& 0&m_{12}\end{array}\right] _{p}, \nn
\eea}}
together with the relation
\be {\small{
 \left[\begin{array}{ccc}\Delta_1-1&\Delta_2-1&\Delta_3\\l_1+1&l_2+1&l_3\\ 0& 0&m_{12}+1\end{array}\right] _p = (I_{12}I_{21})  \left[\begin{array}{ccc}\Delta_1&\Delta_2&\Delta_3\\l_1&l_2&l_3\\0& 0&m_{12}\end{array}\right] _p \,.}}
\ee
Notice that the operators $D_{1,2}$ and $\widetilde D_{1,2}$ relate nearest neighbour classes and the iteration eventually involves all classes at the same time.
The action of the $D_2$ and $\widetilde D_2$ derivatives can be obtained by replacing $1\leftrightarrow 2$, $p\leftrightarrow (\delta-p)$ in the coefficients multiplying the structures and
$p+1\rightarrow p-1$ in the subscripts, and by changing sign on one side of the equation.
Structures with non-vanishing $m_{13}$ and $m_{23}$ are obtained using
 {\small{\bea
&& 4(l_3-m_{13}-m_{23}+\delta-p)  \left[\begin{array}{ccc}\Delta_1&\Delta_2&\Delta_3\\l_1&l_2&l_3\\m_{23}&m_{13}+1&m_{12}\end{array}\right]_p
 - 4(\delta-p) \left[\begin{array}{ccc}\Delta_1&\Delta_2&\Delta_3\\l_1&l_2&l_3\\m_{23}+1&m_{13}&m_{12}\end{array}\right]_{p+1} =  \nn  \\
 && \hspace{2cm} 4 (l_2-\delta+p-m_{23}-m_{12}) \left[\begin{array}{ccc}\Delta_1&\Delta_2&\Delta_3\\l_1&l_2&l_3\\m_{23}&m_{13}&m_{12}+1\end{array}\right]_p
 +D_1 \left[\begin{array}{ccc}\Delta_1&\Delta_2+1&\Delta_3\\l_1-1&l_2&l_3\\m_{23}&m_{13}&m_{12}\end{array}\right] _p \\
&& \hspace{2cm}  - \frac 12 (2m_{12}-2m_{13}+\Delta_2-\Delta_1-\Delta_3 -l_1-l_2+l_3+2\delta-2p+2)  \left[\begin{array}{ccc}\Delta_1&\Delta_2&\Delta_3\\l_1&l_2&l_3\\m_{23}&m_{13}&m_{12}\end{array}\right]_p \nn
\eea}}
together with the corresponding relation with $1\leftrightarrow 2$ and $p\rightarrow p+1$.
All the structures (\ref{deltaGenExpST}) are hence derivable from $\delta+1$ seeds by acting with the operators $D_{1,2}$, $\widetilde D_{1,2}$ and $(I_{12}I_{21})$.
The seeds, on the other hand, are all related by means of the following relation:
\be
(\delta-p)^2\left[\begin{array}{ccc}\Delta_1&\Delta_2&\Delta_3\\ p+1 &\delta-p-1 &l_3\\0&0&0\end{array}\right] _{p+1}
=R  \left[\begin{array}{ccc}\Delta_1+1&\Delta_2+1&\Delta_3\\ p &\delta-p &l_3\\0&0&0\end{array}\right] _{p} \,,
\ee
where
\be
R\equiv -\frac 12 \bar d_2 d_2 \,.
\ee
We conclude that, starting from the single seed correlator with $p=0$,
\begin{equation}{\small{
\left[\begin{array}{ccc}\Delta_1&\Delta_2&\Delta_3\\0&\delta &l_3\\0&0&0\end{array}\right] _0 \equiv  \Big(\frac{K_{1,23} I_{23}}{X_{23}}\Big)^{\delta}
 \left[\begin{array}{ccc}\Delta_1&\Delta_2&\Delta_3\\0& 0 &l_3\\0&0&0\end{array}\right] \,,}}
\label{deltaGenExpSTseed}
\end{equation}
namely the three-point function of a scalar, a spin $\delta$ traceless symmetric operator and the mixed symmetry operator with spin
$(l_3+2\delta,l_3)$,  we can obtain all tensor structures of higher spin correlators.

Let us now see how the constraint on $m_{12}$ in eq.(\ref{j123AO}) arises. When $p\neq 0, \delta$, namely when both $K_1$ and $K_2$ structures appear at the same time, combining eqs.(\ref{JK12}), the following relation is shown to hold:  {\small{
\bea 
&& \left[\begin{array}{ccc}\Delta_1&\Delta_2&\Delta_3\\l_1&l_2&l_3\\m_{23}&m_{13}&m_{12}\!+\!1\end{array}\right] _p  = -\frac 14  \left[\begin{array}{ccc}\Delta_1&\Delta_2&\Delta_3\\l_1&l_2&l_3\\m_{23}&m_{13}&m_{12}\end{array}\right]_p-
 \left[\begin{array}{ccc}\Delta_1&\Delta_2&\Delta_3\\l_1&l_2&l_3\\m_{23}&m_{13}\!+\!1&m_{12}\end{array}\right]_p-
 \left[\begin{array}{ccc}\Delta_1&\Delta_2&\Delta_3\\l_1&l_2&l_3\\m_{23}\!+\!1&m_{13}&m_{12}\end{array}\right]_p \nn \\
 &&\hspace{1.5cm} -8
  \left[\begin{array}{ccc}\Delta_1&\Delta_2&\Delta_3\\l_1&l_2&l_3\\m_{23}\!+\!1&m_{13}\!+\!1&m_{12}\end{array}\right]_{p}+
   \left[\begin{array}{ccc}\Delta_1&\Delta_2&\Delta_3\\l_1&l_2&l_3\\m_{23}&m_{13}\!+\!1&m_{12}\end{array}\right]_{p-1}+4
    \left[\begin{array}{ccc}\Delta_1&\Delta_2&\Delta_3\\l_1&l_2&l_3\\m_{23}&m_{13}\!+\!2&m_{12}\end{array}\right]_{p-1}\nn  \\
   && \hspace{1.5cm} +
     \left[\begin{array}{ccc}\Delta_1&\Delta_2&\Delta_3\\l_1&l_2&l_3\\m_{23}\!+\!1&m_{13}&m_{12}\end{array}\right]_{p+1}+4
      \left[\begin{array}{ccc}\Delta_1&\Delta_2&\Delta_3\\l_1&l_2&l_3\\m_{23}\!+\!2&m_{13}&m_{12}\end{array}\right]_{p+1}\,. 
 \label{deltaGenRelInd} 
\eea}}
Using it iteratively, we can reduce all structures with $p\neq 0, \delta$ to those with $m_{12}=0$ and with $p=0,\delta$, any $m_{12}$.\footnote{One has to recall the range of the parameters 
(\ref{j123AO}), otherwise it might seem that  non-existant  structures can be obtained from eq.(\ref{deltaGenRelInd}).}
This proves the validity of eq.(\ref{deltaGenExpST}). As a further check, we have verified that the number of tensor structures obtained from eq.(\ref{deltaGenExpST})
agrees with those found from eq.(3.38) of ref.\cite{Elkhidir:2014woa}.

\subsubsection{Recursion Relations for general $l_3$}
\label{subsubsec:smalll3}

The tensor structures of correlators with  $l_3< l_1+l_2-\delta$ cannot all be reduced in the form (\ref{deltaGenExpST}), because we are no longer ensured to have enough $J_{3,12}$ invariants
to remove all the $K_{3,12}$'s by means of eq.(\ref{JK3}). In this case the most general tensor structure reads
{\small{
\begin{equation}
\left[\begin{array}{ccc}\Delta_1&\Delta_2&\Delta_3\\l_1&l_2&l_3\\m_{23}&m_{13}&m_{12}\end{array}\right] _{p,q} \!\!\! \equiv\eta  \Big(\frac{K_{1,23} I_{23}}{X_{23}}\Big)^{\delta-p}
 \Big(\frac{K_{2,31} I_{13}}{X_{13}}\Big)^{q} \Big(\frac{K_{3,12} I_{13} I_{23}}{\sqrt{X_{12}X_{13}X_{23}}}\Big)^{p-q}
 \left[\begin{array}{ccc}\Delta_1&\Delta_2&\Delta_3\\l_1-p &l_2-\delta+q &l_3\\m_{23}&m_{13}&\widetilde m_{12}\end{array}\right] \,,
\label{deltaGenExpSTlowl3}
\end{equation}}}
with  $p=0,\ldots,\delta$, $q=0,\ldots,\delta$, $p-q\geq 0$ and
\begin{equation}\begin{aligned}\label{j123AOlowl}
j_1 & = l_1-p - \widetilde m_{12}- m_{13} \geq 0\,, \\
 j_2 & =l_2-\delta+q -\widetilde m_{12}-m_{23}\geq 0 \,,\\
j_3 & =l_3 -m_{13}-m_{23} \geq 0 \,,
\end{aligned}\hspace{1cm}
\begin{aligned}\
\widetilde m_{12} &  = \left\{\begin{array}{l}m_{12} \,\,\,\,\;\;\; if \;\;\;\;\;q=0 \,\,\,or \,\,\, p= \delta \\ 0 \,\,\,\,\,\,\,\,\, \,\;\;otherwise \end{array}\right.   \\  
\eta &   = \left\{\begin{array}{l}0  \,\,\,\,\;\;\; if \;\;\;\; j_3>0 \,\,\,and \,\,\, p\neq q\\ 1 \,\,\, \;\;\;otherwise \end{array}\right.  \,.
\end{aligned}\end{equation}
The parameter $\eta$ in eq.(\ref{j123AOlowl}) is necessary because the tensor structures involving $K_{3,12}$ (i.e. those with $p\neq q$) are independent only when $j_3=0$, namely
when the traceless symmetric structure does not contain any $J_{3,12}$ invariant.
All the tensor structures (\ref{deltaGenExpSTlowl3}) can be reached starting from the single seed with $p=0$, $q=0$, $l_1=0$, $l_2=\delta$ and $m_{ij}=0$.
The analysis follows quite closely the one made for $l_3\geq  l_1+l_2-\delta$, although  it is slightly more involved.
As before, it is convenient to first construct a redundant basis where $m_{12}\neq 0$ for any $p,q$ and we neglect the factor $\eta$ above, and impose only later  the relations that leads to the independent basis (\ref{deltaGenExpSTlowl3}). We start from the structures with $p=q$, which are the same as those in eq.(\ref{deltaGenExpST}): first construct the structures with $m_{13}=m_{23}=0$ by applying iteratively the operators $D_{1,2}+\widetilde D_{1,2}$, and then apply $D_1$ and $D_2$ to get the structures with non-vanishing $m_{13}$ and $m_{23}$. Structures with $p\neq q$ appear when acting with $D_1$ and $D_2$. We have:
 {\small{\bea
&& D_1 \left[\begin{array}{ccc}\Delta_1&\Delta_2+1&\Delta_3\\l_1-1&l_2&l_3\\m_{23}&m_{13}&m_{12}\end{array}\right] _{p,p} = 2(\delta-p) \left[\begin{array}{ccc}\Delta_1&\Delta_2&\Delta_3\\l_1&l_2&l_3\\m_{23}&m_{13}&m_{12}\end{array}\right] _{p+1,p} \label{D1lowl3} \\
&& -4(l_2+p-\delta-m_{12}-m_{23})  \left[\begin{array}{ccc}\Delta_1&\Delta_2&\Delta_3\\l_1&l_2&l_3\\m_{23}&m_{13}&m_{12}+1\end{array}\right] _{p,p}
 +4(l_3-m_{13}-m_{23}) \left[\begin{array}{ccc}\Delta_1&\Delta_2&\Delta_3\\l_1&l_2&l_3\\m_{23}&m_{13}+1&m_{12}\end{array}\right] _{p,p} \nn \\
 && \hspace{2cm}  + \frac 12 \Big(2m_{12}-2m_{13}+\Delta_2-\Delta_1-\Delta_3 -l_1-l_2+l_3+2(\delta-p+1)\Big)  \left[\begin{array}{ccc}\Delta_1&\Delta_2&\Delta_3\\l_1&l_2&l_3\\m_{23}&m_{13}&m_{12}\end{array}\right]_{p,p} \nn  \,.
\eea}}
The action of $D_2$ is obtained by exchanging $1\leftrightarrow 2$ and $\delta-p\leftrightarrow q$ in the coefficients multiplying the structures
and replacing the subscript $(p+1,p)$ with $(p,p-1)$.
For $m_{13}+m_{23}<l_3$ the first term in eq.(\ref{D1lowl3}) is redundant and can be expressed in terms of the known structures with $p=q$.
An irreducible structure is produced only when we reach the maximum allowed value $m_{13}+m_{23}=l_3$, in which case the third term in eq.(\ref{D1lowl3}) vanishes
and we can use the equation to get the irreducible structures with $p\neq q$.
Summarizing, all tensor structures can be obtained starting from a single seed upon the action of the operators $D_{1,2}$, $(D_{1,2}+\widetilde D_{1,2})$, $I_{12} I_{21}$ and $R$.

\subsubsection{Differential Basis}
\label{subsubsec:DBAnti}

A differential basis  that is well defined for any value of $l_1$, $l_2$, $l_3$ and $\delta$ is
\begin{equation}\begin{aligned}
 {\small{\left\{\begin{array}{ccc}\Delta_1&\Delta_2&\Delta_3\\l_1&l_2&l_3\\m_{23}&m_{13}&m_{12}\end{array}\right\}_{p,q} }}& = \eta\,
(I_{12} I_{21})^{\widetilde m_{12}} D_1^{m_{13}+p-q} D_2^{m_{23}}\sum_{n_1=0}^{j_1}\Big(
\begin{array}{c} j_1 \\ n_1\end{array}\Big) D_1^{n_1} \widetilde D_1^{j_1-n_1}
\sum_{n_2=0}^{j_2}\Big(
\begin{array}{c} j_2 \\ n_2\end{array}\Big) \\ &  D_2^{n_2} \widetilde D_2^{j_2-n_2}
 R^q
{\small{\left[\begin{array}{ccc}\Delta_1^\prime&\Delta_2^\prime &\Delta_3\\0 & \delta &l_3\\ 0 & 0 & 0 \end{array}\right]_0 } },
\label{DBevenNew}
\end{aligned}
\end{equation}
where  $\Delta_1^{\prime} =  \Delta_1 + l_1+m_{23}-m_{13}+n_2-n_1-p +q$, $\Delta_2^{\prime} =  \Delta_2 + l_2+m_{13}-m_{23}+n_1-n_2+2q-\delta$, and all parameters are defined as in eq.(\ref{j123AOlowl}).
The recursion relations found above have shown that the differential basis (\ref{DBevenNew}) is  complete. One can also check that its dimensionality matches the one of the ordinary basis for any  $l_1$, $l_2$, $l_3$ and $\delta$. Like in the purely traceless symmetric case,  the specific choice of operators made in eq.(\ref{DBevenNew}) seems to be enough to ensure that the determinant of the transformation matrix is non-vanishing regardless of the choice of $\Delta_1$ and $\Delta_2$. We have explicitly checked this result up to $l_1=l_2=2$, for any $l_3$. The transformation matrix is always of maximal rank, except for the case $l_3=0$ and $\Delta_3=2$, which saturates the unitarity bound for $\delta=1$. Luckily enough, this case is quite trivial, being associated to the
exchange of a free $(2,0)$ self-dual tensor \cite{Siegel:1988gd} (see footnote \ref{foot:1}). The specific ordering of the differential operators is a choice motivated by the form of the recursion relations, as before, and different orderings can be trivially related by using the commutators defined in eq.(\ref{commutators}).

\section{Computation of Four-Point Functions}

We have shown in section \ref{sec:cpw} how relations between three-point functions lead to relations between CPW.
The latter are parametrized by 4-point, rather than 3-point, function tensor structures, so in order to make further progress it is important to classify four-point functions. It should be clear that even when acting on scalar quantities, tensor structures belonging to the class of 4-point functions are generated.  For example $\widetilde D_{1}U=-UJ_{1,24}$.
We postpone to another work a general classification, yet we want to show in the following subsection a preliminary analysis, enough to study the  four fermion correlator  example in subsection \ref{subsec:4fer}.

\subsection{Tensor Structures of Four-Point Functions}

In 6D, the index-free uplift of the four-point function (\ref{Gen4pt})  reads
\begin{equation}
\langle O_1O_2O_3O_4 \rangle=\mathcal{K}_4\;\sum_{n=1}^{N_{4}}g_n(U,V) \; \mathcal{T}^n(S_1,\bar{S}_1,..,S_4,\bar{S}_4),
\end{equation}
where $\mathcal{T}^n$ are the 6D uplifts of the tensor structures  appearing in eq.(\ref{Gen4pt}). The 6D kinematic factor $\mathcal{K}_4$  and the conformally invariant cross ratios $(U,V)$ are obtained from their 4D counterparts by the replacement $x_{ij}^2\rightarrow X_{ij}$ in eqs.(\ref{eq:4kinematic}) and (\ref{uv4d}).

The tensor structures $\mathcal{T}^n$ are formed from the three-point invariants~(\ref{eq:invar1})-(\ref{eq:invar4}) (where $i,j,k$ now run from 1 to 4) and the following new ones:
\begin{align}
\label{eq:invar5}
J_{ij,kl}           &\equiv N_{kl}\, \bar S_i \mathbf{X}_k \overline{\mathbf{X}}_l S_j  \,, \\
\label{eq:invar6}
K_{i,jkl}            &\equiv N_{jkl}\,  S_i \overline{\mathbf{X}}_j \mathbf{X}_k \overline{\mathbf{X}}_l S_i \,, \\
\label{eq:invar7}
\overline{K}_{i,jkl} &\equiv N_{jkl}\, \bar S_i \mathbf{X}_j \overline{\mathbf{X}}_k \mathbf{X}_l \bar S_i \,,
\end{align}
where $i\neq j\neq k\neq l=1,2,3,4$; $K_{i,jkl}$ and $\overline{K}_{i,jkl}$ are totally anti-symmetric in the last three indices and the normalization factor is given by
\begin{equation}
N_{jkl}\equiv \frac{1}{\sqrt{X_{jk}X_{kl}X_{lj}}}.
\end{equation}
The invariants $J_{ij,kl}$ satisfy the relations  $J_{ij,kl}=-J_{ij,lk}+2I_{ij}$.
Given that, and the 4D parity transformations $K_{i,jkl} \stackrel{{\cal P}}{\longleftrightarrow} \overline{K}_{i,jkl}$ and $J_{ij,kl} \stackrel{{\cal P}}{\longleftrightarrow}-J_{ji,lk}$, a convenient choice of index ordering in $J_{ij,kl}$ is
$(i<j,\,k<l)$ and $(i>j,\,k>l)$. Two other invariants $H\equiv  S_1  S_2 S_3 S_4$ and $\bar H\equiv  \bar S_1  \bar S_2 \bar S_3 \bar S_4$ formed by using the epsilon $SU(2,2)$ symbols, are redundant. For instance, one has $X_{12}H=K_{2,14}K_{1,23}-K_{1,24}K_{2,13}$.

Any four-point function can be expressed as a sum of products of the invariants (\ref{eq:invar1})-(\ref{eq:invar4}) and (\ref{eq:invar5})-(\ref{eq:invar7}). However, not every product is independent, due to several relations between them. Leaving to a future work the search of all possible relations, we report in Appendix~\ref{app:relations}  a small subset of them. Having a general classification of 4-point tensor structures is crucial to bootstrap a four-point function with non-zero external spins. 
When we equate correlators in different channels, we have to identify all the factors in front of the same tensor structure, thus
it is important to have a common basis of independent tensor structures.

\subsection{Counting 4-Point Function Structures}

In absence of a general classification of 4-point functions, we cannot directly count the number $N_4$ of their tensor structures.
However, as we already emphasized in ref.\cite{Elkhidir:2014woa}, the knowledge of 3-point functions and the OPE should be enough to infer $N_4$ by means of eq.(\ref{N4N3}).
In this subsection we show how to use eq.(\ref{N4N3}) to determine $N_4$, in particular when parity and permutation symmetries are imposed.

If the external operators are traceless symmetric, the CPW can be divided in parity even and odd. This is clear when the exchanged operator is also traceless symmetric:
$W^{(p,q)}_{O(l,l)+} \stackrel{{\cal P}}{\longrightarrow}W^{(p,q)}_{O(l,l)+}$ if the 3-point structures $p$ and $q$ are both parity even or both parity odd,
$W^{(p,q)}_{O(l,l)-} \stackrel{{\cal P}}{\longrightarrow}-W^{(p,q)}_{O(l,l)-}$ if only one of the structures $p$ or $q$ is parity odd.
For  mixed symmetry exchanged operators $O^{l+2\delta,l}$ or $O^{l,l+2\delta}$, we have $W^{(p,q)}_{O(l+2\delta,\,l)} \stackrel{{\cal P}}{\longrightarrow}W^{(p,q)}_{O(l,\,l+2\delta)}$, so that
$W^{(p,q)}_{Or+}=W^{(p,q)}_{Or}+W^{(p,q)}_{O\bar r}$  is parity even and 
$W^{(p,q)}_{Or-}=W^{(p,q)}_{Or}-W^{(p,q)}_{O\bar r}$ is parity odd.
If parity is conserved, only parity even or odd CPW survive, according to the parity transformation of the external operators. The number of parity even and parity odd 4-point tensor structures are 
\begin{equation}\begin{aligned}\label{n4Parity}
N_{4+}&=N^{12}_{3(l,l)+}N^{34}_{3(l,l)+}+N^{12}_{3(l,l)-}N^{34}_{3(l,l)-}+\sum_{r\neq(l,l)}\frac{1}{2}N^{12}_{3r}N^{34}_{3\bar r} \,, \\
N_{4-}&=N^{12}_{3(l,l)-}N^{34}_{3(l,l)+}+N^{12}_{3(l,l)+}N^{34}_{3(l,l)-}+\sum_{r\neq(l,l)}\frac{1}{2}N^{12}_{3r}N^{34}_{3\bar r}\,.
\end{aligned}\end{equation}
The numbers $N_{4+}$ and $N_{4-}$ in eq.(\ref{n4Parity}) are always integers, because in the sum over $r$ one has to consider separately $r=(l,\bar l)$ and $r=(\bar l,l)$,\footnote{Recall that $r$ is not an infinite sum over all possible spins, but a finite sum over the different classes of representations, see eq.(\ref{sch4pt}) and text below.} and  which give an equal contribution that compensates for the factor 1/2.

When some of the external operators are equal,  permutation symmetry should be imposed. We consider here only the permutations $1\leftrightarrow3$, $2\leftrightarrow4$ and
$1\leftrightarrow 2$, $3\leftrightarrow4$ that leave $U$ and $V$ invariant and simply give rise to a reduced number of tensor structures.
Other permutations would give relations among the various functions $g_n(U,V)$ evaluated at different values of their argument.
If  $O_1=O_3$, $O_2=O_4$, the CPW in the s-channel transforms as follows under the permutation $1\leftrightarrow3$, $2\leftrightarrow4$:
$W^{(p,q)}_{Or}\xrightarrow{per} W^{(q,p)}_{O\bar r}$. We then have $W^{(p,q)}_{O(l,l)+}=W^{(q,p)}_{O(l,l)+}$, $W^{(p,q)}_{O(l,l)-}=W^{(q,p)}_{O(l,l)-}$, $W^{(p,q)}_{Or+}=W^{(q,p)}_{Or+}$, $W^{(p,q)}_{Or-}=-W^{(q,p)}_{Or-}$. The number of parity even and parity odd 4-point tensor structures in this case is
\begin{equation}\begin{aligned}\label{n4ParityPermutation}
N_{4+}^{per}&=\frac12N^{12}_{3(l,l)+}(N^{34}_{3(l,l)+}+1)+\frac12N^{12}_{3(l,l)-}(N^{34}_{3(l,l)-}+1)+\sum_{r\neq(l,l)}\frac{1}{4}N^{12}_{3r}(N^{12}_{3r}+1)\,, \\
N_{4-}^{per}&=N^{12}_{3(l,l)-}N^{12}_{3(l,l)+}+\sum_{r\neq(l,l)}\frac{1}{4}N^{12}_{3r}(N^{12}_{3r}-1) \,,
\end{aligned}\end{equation}
where again in the sum over $r$ one has to consider separately $r=(l,\bar l)$ and $r=(\bar l,l)$.
If  $O_1=O_2$, $O_3=O_4$, the permutation $1\leftrightarrow2$, $3\leftrightarrow4$ reduces the number of tensor structures of the CPW in the s-channel,  $N^{12}_3\rightarrow  N^{1=2}_3\leq N^{12}_3$ and $N^{34}_3\rightarrow N^{3=4}_3\leq N^{12}_3$.  Conservation of external operators has a similar effect.

\subsection{Relation between ``Seed" Conformal Partial Waves}

Using the results of the last section, we can compute the CPW associated to the exchange of arbitrary 
operators with external traceless symmetric fields, in terms of a set of seed CPW, schematically denoted by $W_{{\cal O}^{l+2\delta,l}}^{(p,q)}(l_1,l_2,l_3,l_4)$.
We have
\be
W_{O^{l+2\delta,l}}^{(p,q)}(l_1,l_2,l_3,l_4) = D_{(12)}^{(p)} D_{(34)}^{(q)} W_{O^{l+2\delta,l}}(0,\delta,0,\delta)\,, 
\label{Wl1234}
\ee
where $D_{12}^{(p)}$ schematically denotes the action of the 
differential operators reported in the last section, and $D_{34}^{(q)}$ are the same operators for the fields at $X_3$ and $X_4$, obtained by replacing $1\rightarrow 3$, $2\rightarrow 4$ everywhere in eqs.(\ref{DDtilde})-(\ref{Dtilde12}) and (\ref{nablaS}).
For simplicity we do not report the dependence of $W$ on $U,V$, and on the scaling dimensions 
of the external and exchanged operators. The seed CPW are the simplest among the ones appearing in correlators of traceless symmetric tensors, but they are {\it not} the simplest in general. These will be the CPW arising from the  four-point functions with the {\it lowest} number of tensor structures with a non-vanishing contribution of the field $O^{l+2\delta,l}$ in some of the OPE channels. 
Such minimal four-point functions are\footnote{Instead of eq.(\ref{4ptanti}) one could also use 4-point functions with two scalars and two $O^{(0,2\delta)}$ fields  or two scalars and two $O^{(2\delta,0)}$ fields. Both have the same number $2\delta+1$ of tensor structures as the correlator (\ref{4ptanti}).}
\be
\langle O^{(0,0)}(X_1)  O^{(2\delta,0)}(X_2)   O^{(0,0)}(X_3)  O^{(0,2\delta)}(X_4) \rangle  = \mathcal{K}_4  \sum_{n=0}^{2\delta} g_n(U,V)  I_{42}^n J_{42,31}^{2\delta-n}\,,
\label{4ptanti}
\ee
with just
 \be
N_4^{seed}(\delta)=2\delta+1
\ee
tensor structures. In the s-channel (12-34) operators $O^{l+n,l}$, with $-2\delta\leq n\leq 2\delta$, are exchanged. 
We denote by $W_{seed}(\delta)$ and $\overline{W}_{seed}(\delta)$ the single CPW associated to the exchange of the fields $O^{l+2\delta,l}$ and  $O^{l,l+2\delta}$ in the four-point function (\ref{4ptanti}).
They are parametrized in terms of $2\delta+1$ conformal blocks as follows (${\cal G}_0^{(0)}= \overline{{\cal G}}_0^{(0)}$):
\bea
W_{seed}(\delta)& = &  \mathcal{K}_4  \sum_{n=0}^{2\delta} {\cal G}_n^{(\delta)}(U,V)  I_{42}^n J_{42,31}^{2\delta-n}\,,\nn \\
\overline W_{seed}(\delta)& = & \mathcal{K}_4  \sum_{n=0}^{2\delta} \overline{\cal G}_n^{(\delta)}(U,V)  I_{42}^n J_{42,31}^{2\delta-n}\,.
\label{W2deltaExp}
\eea
In contrast, the number of tensor structures in $\langle O^{(0,0)}(X_1)  O^{(\delta,\delta)}(X_2) O^{(0,0)}(X_3)   O^{(\delta,\delta)}(X_4) \rangle$  grows rapidly with $\delta$.  Denoting it by $\widetilde N_4(\delta)$ we have, using eq.(6.6) of ref.\cite{Elkhidir:2014woa}:
 \be
 \widetilde N_4(\delta) = \frac 13 \Big(2 \delta^3+6\delta^2+7\delta+3\Big)\,.
 \ee
It is important to stress that a significant simplification occurs in using seed CPW even when there is no need to reduce their number, i.e. $p=q=1$.
For instance, consider the correlator of four traceless symmetric spin 2 tensors. The CPW $W_{O^{l+8,l}}(2,2,2,2)$ is unique, yet it contains 1107 conformal blocks
(one for each tensor structure allowed in this correlator), to be contrasted to the 85 present in $W_{O^{l+8,l}}(0,4,0,4)$ and the 9 in $W_{seed}(4)$! We need to relate $\langle  O^{(0,0)}(X_1) O^{(2\delta,0)}(X_2) O^{(l+2\delta,l)}(X_3)\rangle$ and $\langle  O^{(0,0)}(X_1) O^{(\delta,\delta)}(X_2) O^{(l+2\delta,l)}(X_3)\rangle$ in order to be able to use the results of section \ref{sec:DBTSO} together with $W_{seed}(\delta)$. As explained at the end of Section \ref{sec:OB}, there is no combination of first-order operators which can do this job and one is forced to use the operator~(\ref{nablaS}):
\begin{equation}
\langle O^{(0,0)}_{\Delta_1}(X_1) O^{(\delta,\delta)}_{\Delta_2}(X_2)   O_{\Delta}^{(l,\,l+2\delta)}(X)\rangle_1=   \Big(\prod_{n=1}^\delta c_n\Big)
(\bar{d}_{1} \nabla_{12}  \widetilde D_{1})^\delta
  \langle O^{(0,0)}_{\Delta_1+\delta}(X_1)   O^{(2\delta,0)}_{\Delta_2}(X_2)   O^{(l,\,l+2\delta)}_{\Delta}(X)\rangle_1 \,,
\label{Deltad1t}
\end{equation}
where\footnote{Notice that the scalings dimension $\Delta_1$ and $\Delta_2$ in eq.(\ref{cn}) do not exactly correspond in general to those of the external operators, but should be identified with 
$\Delta_1^\prime$ and $\Delta_2^\prime$ in eq.(\ref{DBevenNew}). It might happen that the coefficient $c_n$ vanishes for some values of $\Delta_1$ and $\Delta_2$. 
As we already pointed out, there is some redundancy that allows us to choose a different set of operators. Whenever this coefficient vanishes, we can choose a different operator, e.g. $\widetilde D_1\rightarrow  D_1$.}
\be
c_n^{-1} =  2(1-n+2\delta)\Big(2(n+1)+\delta+l+\Delta_1-\Delta_2+\Delta\Big)\,.
\label{cn}
\ee
Equation (\ref{Deltad1t}) implies the following relation between the two CPW:
\be
W_{O^{l+2\delta,l}}(0,\delta,0,\delta) = \Big(\prod_{n=1}^\delta c_n^{12} c_n^{34}\Big)
\Big(\nabla_{43} d_{3} \widetilde D_{3}\Big)^\delta
\Big(\nabla_{12} \bar{d}_{1} \widetilde D_{1}\Big)^\delta  W_{seed}(\delta)  \,,
\label{WtoW}
\ee
where $c_n^{12}=c_n$ in eq.(\ref{cn}), $c_n^{34}$ is obtained from $c_n$ by exchanging $1\rightarrow 3, 2\rightarrow 4$ and the scaling dimensions of the corresponding external operators are related as indicated in eq.(\ref{Deltad1t}).

Summarizing, the whole highly non-trivial problem of computing $W_{O^{l+2\delta,l}}^{(p,q)}(l_1,l_2,l_3,l_4)$ has been reduced to the computation of the $2\times (2\delta+1)$ conformal blocks ${\cal G}_n^{(\delta)}(U,V)$  and $ \overline{\cal G}_n^{(\delta)}(U,V)$ entering eq.(\ref{W2deltaExp}).  Once they are known, one can use eqs.(\ref{WtoW}) and (\ref{Wl1234})  to finally  reconstruct $W_{O^{l+2\delta,l}}^{(p,q)}(l_1,l_2,l_3,l_4)$.

\section{Examples}

In this section we would like to elucidate various aspects of our construction. In the subsection~\ref{subsec:4fer} we give an example in which we deconstruct a correlation function of four fermions. We leave the domain of traceless symmetric external operators to show the generality of our formalism. It might also have some relevance in phenomenological applications beyond the Standard Model \cite{Caracciolo:2014cxa}. In the subsection~\ref{subsec:ConservedOperators} we consider the special cases of correlators with four conserved identical operators, like spin 1 currents and energy momentum tensors, whose relevance is obvious. There we will just outline the main steps focusing on the implications of current conservations and permutation symmetry in our deconstruction process. 

\subsection{Four Fermions Correlator}
\label{subsec:4fer}

Our goal here is to deconstruct the CPW in the s-channel associated to the 
four fermion correlator
\be
\langle \bar\psi^{\dot \alpha}(x_1)\psi_\beta (x_2)\chi_\gamma(x_3) \bar\chi^{\dot \delta}(x_4) \rangle \,.
\label{4ferm4D}
\ee
For simplicity, we take  $\bar \psi$ and $\bar \chi$ to be conjugate fields of $\psi$ and $\chi$, respectively, so that we have only two different scaling dimensions, $\Delta_\psi$ and $\Delta_\chi$.
Parity invariance is however not imposed in the underlying CFT. The correlator (\ref{4ferm4D}) admits six different tensor structures. 
An independent basis of tensor structures for the 6D uplift of  eq. (\ref{4ferm4D}) can be found using the relation~(\ref{eq:rel8}).
A possible choice is 
\bea
\label{eq:4-pf}
&& \!\!\!\!\! \langle \Psi(X_1,\,\bar S_1)\,\bar{\Psi}(X_2,\, S_2)\,\bar{\mathcal{X}}(X_3,\,S_3)\,\mathcal{X}(X_4,\,\bar S_4) \rangle=\frac{1}{X_{12}^{\Delta_\psi+\tfrac{1}{2}}X_{34}^{\Delta_\chi+\tfrac{1}{2}}}
\bigg(g_1(U,V)I_{12}I_{43}+ \\ 
&& \!\!\!\!\! g_2(U,V)I_{42}I_{13}+g_3(U,V)I_{12}J_{43,21}+
g_4(U,V)I_{42}J_{13,24}+g_5(U,V)I_{43}J_{12,34}+g_6(U,V)I_{13}J_{42,31}\bigg). \nn
\eea
For $l\geq 1$, four CPW $W_{O^{l,l}}^{(p,q)}$ ($p,q=1,2$) are associated to the exchange of traceless symmetric fields, and one for each mixed symmetry field,  $W_{O^{l+2,l}}$ and $W_{O^{l,l+2}}$. Let us start with $W_{O^{l,l}}^{(p,q)}$.
The traceless symmetric CPW are obtained as usual by relating the three point function of two fermions and one $O^{l,l}$ to that of two scalars and one $O^{l,l}$.
This relation requires to use the operator (\ref{nablaS}).
There are two tensor structures for $l\geq 1$:
\bea \label{FerTS}
\langle \Psi(\bar S_1) \bar{\Psi}(S_2) O^{l,l} \rangle_1 & = &  \mathcal{K}  I_{12}J_{0,12}^l =I_{12} \langle \Phi^{\frac 12} \Phi^{\frac 12}  O^{l,l} \rangle_1,  \\
 \langle \Psi(\bar S_1)  \bar{\Psi}(S_2) O^{l,l} \rangle_2 & = & \mathcal{K}  I_{10}I_{02}J_{0,12}^{l-1} =\frac{1}{16l(\Delta-1)} \nabla_{21} \Big( \widetilde D_2 \widetilde D_1  
+\kappa I_{12} \Big) \langle \Phi^{\frac 12}  \Phi^{\frac 12}  O^{l,l} \rangle_1, \nn
\eea
where $\kappa=2\big(4\Delta-(\Delta+l)^2  \big)$, the superscript $n$ in $\Phi$ indicates the shift in the scaling dimensions of the field 
and the operator $O^{l,l}$ is taken at $X_0$.  Plugging eq.(\ref{FerTS}) (and the analogous one for $\mathcal{X}$ and $\bar{\mathcal{X}}$) in eq.(\ref{shadow2}) gives the relation between CPW.
In order to simplify the equations, we report below the CPW in the differential basis, the relation with the ordinary basis being easily determined from eq.(\ref{FerTS}):
\begin{equation}  
\begin{aligned}
W_{O^{l,l}}^{(1,1)} = & I_{12} I_{43} W^{\frac 12,\frac 12,\frac 12,\frac 12}_{seed}(0)\,, \\
W_{O^{l,l}}^{(1,2)} = & I_{12} \nabla_{34} \widetilde D_4 \widetilde D_3 W^{\frac 12,\frac 12,\frac 12,\frac 12}_{seed}(0)  \,, \\
W_{O^{l,l}}^{(2,1)} = & I_{43} \nabla_{21} \widetilde D_2 \widetilde D_1 W^{\frac 12,\frac 12,\frac 12,\frac 12}_{seed}(0)  \,, \\
W_{O^{l,l}}^{(2,2)} = & \nabla_{21} \widetilde D_2 \widetilde D_1 \nabla_{34} \widetilde D_4 \widetilde D_3 W^{\frac 12,\frac 12,\frac 12,\frac 12}_{seed}(0) \,,
 \end{aligned}
\label{CPWferExp}
\end{equation}
where $\widetilde D_3$ and $\widetilde D_4$ are obtained from $\widetilde D_1$ and $\widetilde D_2$ in eq.(\ref{DDtilde}) by replacing $1\rightarrow 3$ and $2\rightarrow 4$ respectively. The superscripts indicate again the shift in the scaling dimensions of the external operators.
As in ref.\cite{Costa:2011dw} the CPW associated to the exchange of traceless symmetric fields is entirely determined in terms of the single known CPW of four scalars $W_{seed}(0)$.
For illustrative purposes, we report here the explicit expressions of  $W_{O^{l,l}}^{(1,2)}$:
\begin{multline}
\mathcal{K}_4^{-1}W_{O^{l,l}}^{(1,2)} = 8I_{12}I_{43}\Bigg(U\big(V-U-2\big)\partial_U + U^2\big(V-U\big)\partial_U^2 + \big(V^2-(2+U)V+1\big)\partial_V +\\ V\big(V^2-(2+U)V+1\big)\partial_V^2 +2UV\big(V-U-1\big)\partial_U\partial_V\Bigg){\cal G}_0^{(0)} \\
+ 4UI_{12}J_{43,21}\Bigg(U\partial_U+U^2\partial_U^2+\big(V-1\big)\partial_V+V\big(V-1\big)\partial_V^2 +2UV\partial_U\partial_V
\Bigg){\cal G}_0^{(0)},
\label{W12Ol}
\end{multline}
where ${\cal G}_0^{(0)}$ are the known scalar conformal blocks \cite{Dolan:2000ut,Dolan:2003hv}.
It is worth noting that the relations~(\ref{eq:rel1})-(\ref{eq:rel8}) have to be used to remove redundant structures and write the above result (\ref{W12Ol}) in the chosen basis~(\ref{eq:4-pf}).
 
The analysis for the mixed symmetry CPW  $W_{O^{l+2,l}}$ and $W_{O^{l,l+2}}$ is simpler.  
The three point function of two fermions and one $O^{l,l+2}$ field has a unique tensor structure, like the one of a scalar and a $(2,0)$ field $F$. One has
\begin{equation}  
\begin{aligned}
\langle \Psi(\bar S_1) \bar\Psi(S_2) O^{l+2,l} \rangle_1 & =   \mathcal{K}  I_{10}K_{1,20} J_{0,12}^l = \frac 14 \bar d_{2} \langle \Phi^{\frac12} F^{\frac12} O^{l+2,l} \rangle_1 \,, \\
\langle \Psi(\bar S_1) \bar\Psi(S_2) O^{l,l+2} \rangle_1 & =   \mathcal{K}  I_{02}\overline K_{2,10} J_{0,12}^l = \frac 12 \bar d_{2} \langle \Phi^{\frac12} F^{\frac12} O^{l,l+2} \rangle_1
\end{aligned}
\label{CPWfer2Exp}
\end{equation}
and similarly for the conjugate $(0,2)$ field $\bar F$.
Using the above relation, modulo an irrelevant constant factor, we get
\begin{equation}  
\begin{aligned}
W_{O^{l+2,l}} = & \;  \bar d_2 d_4 W_{seed}^{\frac12,\frac12,\frac12,\frac12}(1) \,, \\
W_{O^{l,l+2}} =  &\; \bar d_2 d_4 \overline W_{seed}^{\frac12,\frac12,\frac12,\frac12}(1) \,,
 \end{aligned}
\label{CPWfer4Exp}
\end{equation}
where  $W_{seed}(1)$ and $\overline W_{seed}(1)$ are defined in eq.(\ref{W2deltaExp}).
Explicitly, one gets
\begin{equation} 
\begin{aligned}
 \frac{\sqrt{U}}{4}\mathcal{K}_4^{-1}W_{O^{l+2,l}}  =  &  I_{12}I_{43} \Big({\cal G}_2^{(1)} +(V-U-1){\cal G}_1^{(1)} +4U{\cal G}_0^{(1)}\Big) -4U I_{42} I_{13} {\cal G}_1^{(1)} 
 +U I_{12} J_{43,21} {\cal G}_1^{(1)} \\  & -U I_{42} J_{13,24} {\cal G}_2^{(1)}+U I_{43}J_{12,34} {\cal G}_1^{(1)}-4U I_{13}J_{42,31} {\cal G}_0^{(1)}\,.
 \end{aligned}
 \label{CPWfer5Exp} 
\end{equation}
The same applies for $W_{O^{l,l+2}}$ with  ${\cal G}_n^{(1)}\rightarrow \overline  {\cal G}_n^{(1)}$. 
The expression (\ref{CPWfer5Exp}) shows clearly how the six conformal blocks entering $W_{O^{l,l+2}}$ are completely determined in terms of the three ${\cal G}_n^{(1)}$.

\subsection{Conserved Operators}
\label{subsec:ConservedOperators}

In this subsection we outline, omitting some details, the deconstruction of four identical currents and four energy-momentum tensor correlators, which are among the most interesting and universal correlators to consider.  In general,  current conservation relates the coefficients $\lambda_s$ of the three-point function and reduces the number of independent tensor structures.
Since CPW are determined in terms of products of two 3-point functions, the number of  CPW $\widetilde W_{O}$  associated to external conserved operators is reduced with respect to the
one of CPW for non-conserved operators $W_{O}$: 
\begin{equation}
\sum_{p,q=1}^{N_3}\lambda^p_{12O}\lambda^q_{34\bar O} W_{O}^{(p,q)} \longrightarrow \sum_{\tilde p,\tilde q =1}^{\tilde N_3} \lambda^{\tilde p}_{12O}\lambda^{\tilde q} _{34\bar O}
\widetilde W_{O}^{(\tilde p,\tilde q)}\,,
 \end{equation}
where $\tilde N_3\leq N_3$ and 
\be
\widetilde W_{O}^{(\tilde p,\tilde q)}=\sum_{p,q=1}^{N_3} F^{\tilde p p}_{12O}F^{\tilde q q }_{34\bar O} W_{O}^{(p,q)} \,.
\label{CPWtilda}
\ee
The coefficients $F^{\tilde p p}_{12O}$ and $F^{\tilde q q}_{34\bar O}$ depend in general  on the scaling dimension $\Delta$ and spin $l$ of the exchanged operator $O$.
They can be determined by applying the operator defined in eq.(\ref{ConservedD}) to 3-point functions.

\subsubsection{Spin 1 Four-Point Functions}

In any given channel, the exchanged operators are in the $(l,l)$, $(l+2,l)$, $(l,l+2)$, $(l+4,l)$ and $(l,l+4)$ representations. The number of 3-point function tensor structures of these operators with the two external vectors and the total number of four-point function structures is reported in table \ref{tableConservedCurrent}. 
Each CPW can be expanded  in terms of the 70 tensor structures for a total of 4900 scalar conformal blocks as defined in eq.(\ref{WGen}).
Using the differential basis, the $36\times 70=2520$ conformal blocks associated to the traceless symmetric CPW are determined in terms of the single known scalar CPW \cite{Costa:2011dw}. The $16\times 70=1120$ ones associated to ${\cal O}^{l+2,l}$ and ${\cal O}^{l,l+2}$ are all related to the two CPW $W_{seed}(1)$ and $\overline W_{seed}(1)$. Each of them is a function of 3 conformal blocks, see eq.(\ref{W2deltaExp}), for a total of 6 unknown.
Finally, the $2\times 70=140$  conformal blocks associated to ${\cal O}^{l+4,l}$ and ${\cal O}^{l,l+4}$ are expressed in terms of the $5\times 2=10$  conformal blocks coming from the two CPW $W_{seed}(2)$ and $\overline W_{seed}(2)$.

Let us see more closely the constrains coming from permutation symmetry and conservation.
For $l\geq 2$,   the $5_++1_-$ tensor structures of the three-point function  $\langle V_1 V_2 O_{l,l}\rangle$, for distinct non-conserved vectors, reads
\bea
\langle V_1 V_2 O^{l,l}\rangle
&=&\mathcal{K}_3\Big(\lambda_1 I_{23}I_{32}J_{1,23}J_{3,12}+\lambda_2I_{13}I_{31}J_{2,31}J_{3,12}+\lambda_3I_{12}I_{21}J_{3,12}^2 \\
&& +\lambda_4 I_{13}I_{31}I_{23}I_{32}+\lambda_5J_{1,23}J_{2,31}J_{3,12}^2+\lambda_6( I_{21}I_{13}I_{32}+I_{12}I_{23}I_{31} )J_{3,12}\Big)J_{3,12}^{l-2} \,. \nn
\eea
Taking $V_1=V_2$ and applying the conservation condition to the external vectors gives a set of constraints for the OPE coefficients $\lambda_p$. For $\Delta\neq l+4$, we have\footnote{This is the result for generic non-conserved operators $O^{l,l}$.}
\begin{table}
\centering
\begin{tabular}{|c|c|c|c|c|c|c|c|}
\hline
& \multicolumn{2}{|c|}{$O_{l,l}$}&\multicolumn{2}{|c|}{$O_{l+2,l}$}&\multicolumn{2}{|c|}{$O_{l+4,l}$}&\multicolumn{1}{||c|}{$N_4$}\\
$l=$&$2n$&$2n+1$&$2n$&$2n+1$&$2n$&$2n+1$&\multicolumn{1}{||c|}{}\\
\hline
$N^{12}_O$& \multicolumn{2}{|c|}{$5_+ +1_-$}&\multicolumn{2}{|c|}{4}&\multicolumn{2}{|c|}{1}&\multicolumn{1}{||c|}{$43_++27_-$}\\
\hline
$N^{1=2}_O$&$4_+$&$1_+ +1_-$&2&2&1&0&\multicolumn{1}{||c|}{$19_++3_-$}\\
\hline
$N^{1=2}_O$&\multirow{2}{*}{$2_+$}&\multirow{2}{*}{$1_-$}&\multirow{2}{*}{1}&\multirow{2}{*}{1}&\multirow{2}{*}{1}&\multirow{2}{*}{0}&\multicolumn{1}{||c|}{\multirow{2}{*}{$7_+$}}\\ 
conserved&&&&&&&\multicolumn{1}{||c|}{}\\
\hline
\end{tabular}
\caption{Number of independent tensor structures in the 3-point function $\langle V_1 V_2 O^{l,\bar l}\rangle$ when min$(l,\bar l)\geq2-\delta$. In the last column we report $N_4$ as computed using eqs.(\ref{n4Parity}) and (\ref{n4ParityPermutation}) for general four spin 1, identical four spin 1 and identical conserved currents respectively. Subscripts + and - refers to parity even and parity odd structures. For conjugate fields we have $N^{12}_{O(l,l+\delta)}=N^{12}_{O(l+\delta,l)}$.}
\label{tableConservedCurrent}
\end{table}
\begin{equation}\begin{aligned}
F_{12O}^{\tilde p p}(\Delta ,l=2n )=& \begin{pmatrix}1&1&c&a&0&0\\
-\frac12&-\frac12&-\frac12&b&-\frac18&0
\end{pmatrix}, \ \ \ \ 
F_{12O}^{\tilde p p}(\Delta ,l=2n+1 )=\begin{pmatrix}0&0&0&0&0&0\\0&0&0&0&0&1
\end{pmatrix} \,,
\end{aligned}
\label{FlambdaCon}
\end{equation}
with 
\be
a=8\frac{\Delta(\Delta+l+9)-
l(l+8)}{(\Delta-l-4)(\Delta+l)}\,, \ \
b=-4\frac{(\Delta-l-2)}{\Delta-l-4}\,, \ \ 
c=\frac{-\Delta+l+6}{\Delta+l} \,,
\ee
where $F_{12O}^{\tilde p p}$ are the coefficients entering eq.(\ref{CPWtilda}). The number of independent tensor structures is reduced from 6 to $2_+$ when $l$ is even and from 6 to  $1_-$ when $l$ is odd, as indicated in the table \ref{tableConservedCurrent}. When $\Delta = l+4$, eq.(\ref{FlambdaCon}) is modified, but the number of constraints remains the same.
The 3-point function structures obtained, after conservation and permutation is imposed, differ between even and odd  $l$. Therefore,  we need to separately consider the even and odd $l$ contributions when computing $N_4$ using eq.(\ref{n4ParityPermutation}). For four identical conserved currents, $N_4=7_+$, as indicated in table \ref{tableConservedCurrent}, and agrees with what found in ref.\cite{Dymarsky:2013wla}.

\subsubsection{Spin 2 Four-Point Functions}

The exchanged operators can be in the representations $(l+2\delta,l)$ and $(l,l+2\delta)$ where $\delta= 0,1,...,4$. 
The number of tensor structures in the three-point functions of these operators with two external spin 2 tensors is shown in table \ref{tableConservedTensor}. 
We do not list here the number of CPW and conformal blocks  for each representation, which could be easily derived from table \ref{tableConservedTensor}.
In the most general case of four distinct non conserved operators, no parity imposed, one should compute $1107^2\sim 10^6$ conformal blocks, that are reduced to 49 using the differential basis, $W_{seed}(\delta)$ and $\overline W_{seed}(\delta)$.

\begin{table}
\centering
\begin{tabular}{|c|c|c|c|c|c|c|c|c|c|c|c|}
\hline
& \multicolumn{2}{|c|}{$O_{l,l}$}&\multicolumn{2}{|c|}{$O_{l+2,l}$}&\multicolumn{2}{|c|}{$O_{l+4,l}$}&\multicolumn{2}{|c|}{$O_{l+6,l}$}&\multicolumn{2}{|c|}{$O_{l+8,l}$}&\multicolumn{1}{||c|}{$N_4$}\\
$l=$&$2n$&\!$2n\!+\!1$\!&$2n$&\!$2n\!+\!1$\!&$2n$&\!$2n\!+\!1$\!&$2n$&\!$2n\!+\!1$\!&$2n$&\!$2n\!+\!1$\!&\multicolumn{1}{||c|}{}\\
\hline
$N^{12}_{O}$& \multicolumn{2}{|c|}{$14_+ \!+\!5_-$}&\multicolumn{2}{|c|}{16}&\multicolumn{2}{|c|}{10}&\multicolumn{2}{|c|}{4}&\multicolumn{2}{|c|}{1}&\multicolumn{1}{||c|}{$594_+\!+\!513_-$}\\
\hline
$N^{1=2}_O$&$10_+\!+\! 1_-$&$4_+\!+\! 4_-$&8&8&6&4&2&2&1&0&\multicolumn{1}{||c|}{$186_+\!+\!105_-$}\\
\hline
$N^{1=2}_O$&\multirow{2}{*}{$3_+$}&\multirow{2}{*}{$2_-$}&\multirow{2}{*}{2}&\multirow{2}{*}{2}&\multirow{2}{*}{2}&\multirow{2}{*}{1}&\multirow{2}{*}{1}&\multirow{2}{*}{1}&\multirow{2}{*}{1}&\multirow{2}{*}{0}&\multicolumn{1}{||c|}{\multirow{2}{*}{$22_+\!+\!3_-$}}\\
cons.&&&&&&&&&&&\multicolumn{1}{||c|}{}\\
\hline
\end{tabular}
\caption{Number of independent tensor structures in the 3-point function $\langle T_1 T_2 O^{l,\bar l}\rangle$ when min$(l,\bar l)\geq4-\delta$. In the last column we report $N_4$ as computed using eqs.(\ref{n4Parity}) and (\ref{n4ParityPermutation}) for general four spin 2, identical four spin 2 and energy momentum  tensors respectively. Subscripts $+$ and $-$ refers to parity even and parity odd structures. 
For conjugate fields we have $N^{12}_{O(l,l+\delta)}=N^{12}_{O(l+\delta,l)}$.}
\label{tableConservedTensor}
\end{table}

The constraints coming from permutation symmetry and conservation are found as in the spin 1 case, but are more involved and will not be reported.
For four identical spin 2 tensors, namely for four energy momentum tensors, using eq.(\ref{n4ParityPermutation}) one immediately gets $N_4=22_++3_{-}$, as indicated in table \ref{tableConservedCurrent}. The number of parity even structures agrees with what found in ref.\cite{Dymarsky:2013wla}, while to the best of our knowledge the 3 parity odd structures found is a new result.

Notice that even if the number of tensor structures is significantly reduced when conservation is imposed, they are still given by a linear combination of all the tensor structures, 
as indicated in eq.(\ref{CPWtilda}). It might be interesting to see if there exists a formalism that automatically gives a basis of independent tensor structures for conserved operators bypassing eq.(\ref{CPWtilda}) and the use of the much larger basis of allowed structures.

\section{Conclusions}

We have introduced in this paper a set of differential operators, eqs.(\ref{DDtilde}), (\ref{D12}) (\ref{Dbar12}) and (\ref{nablaS}), that enables us to relate different three-point functions in 4D CFTs.
The 6D embedding formalism in twistor space with an index free notation, as introduced in ref.\cite{SimmonsDuffin:2012uy}, and the recent classification of three-point functions in 4D CFTs \cite{Elkhidir:2014woa}  have been crucial to perform this task.
In particular, three-point tensor correlators with different tensor structures can always be related to a three-point function with a single tensor structure.
Particular attention has been devoted to the three point functions of two traceless symmetric and one mixed tensor operator, where explicit independent bases have been provided, eqs.(\ref{deltaGenExpSTlowl3}) and (\ref{DBevenNew}). These results allow us to deconstruct four point tensor correlators,  since we can express the CPW
in terms of a few CPW seeds. We argue that the simplest CPW seeds are those associated to the four point functions of two scalars, one ${\cal O}^{2\delta ,0}$ and one ${\cal O}^{0, 2\delta}$ field, that have only $2\delta+1$ independent tensor structures.

We are now one step closer to bootstrapping tensor correlators in 4D CFTs.
There is of course one important  task to be accomplished: the computation of the seed CPW.
One possibility is to use the shadow formalism as developed in ref.\cite{SimmonsDuffin:2012uy}, 
or  to apply the Casimir operator to the above four point function seeds, hoping that the second order set of partial differential equations for the conformal blocks
is tractable. In order to bootstrap general tensor correlators,  it is also necessary to have a full classification of 4-point functions in terms of $SU(2,2)$ invariants. 
This is a non-trivial task, due to the large number of relations between
the four-point function $SU(2,2)$ invariants. A small subset of them has been reported in the appendix A but many more should be considered for a full classification.
We hope to address these problems in future works.

We believe that universal 4D tensor correlators, such as four energy momentum tensors, might no longer be a dream and are appearing on the horizon!

\section*{Acknowledgments}

We thank Jo$\tilde {\rm a}$o Penedones for useful discussions. The work of M.S. was supported by the ERC Advanced Grant no. 267985 DaMESyFla.

\appendix

%%%%%%%%%%%%%%%%%%%%%%%%%%%%%%%%%%%%%%
%%%%%%%%%%%%%%%%%%%%%%%%%%%%%%%%%%%%%%
\section{Relations between Four-Point Function Invariants}
\label{app:relations}

In this appendix we report a partial list of relations between $SU(2,2)$ invariants entering four-point functions that 
have been used in subsection \ref{subsec:4fer}. 

The first relation is linear in the invariants and reads
\begin{equation}
J_{i,jl}=n_{ijkl}J_{i,kl}+n_{lijk}J_{i,jk}\,,
\end{equation}
where we have defined 
\begin{equation}
n_{ijkl}\equiv\frac{X_{ij}X_{kl}}{X_{ik}X_{jl}}.
\end{equation}
The $7$ relations below allow to eliminate completely products of the form $K_{i,jk}\overline K_{l,mn}$
\begin{align}
\label{eq:rel1}K_{i,jk}\overline K_{i,jk}&=\frac{1}{2} J_{j,ik} J_{k,ij}-2I_{jk} I_{kj} \,, \\
\label{eq:rel2}K_{i,jk}\overline K_{l,jk} &=\sqrt{n_{ijkl}}\Big(n_{iljk}I_{jk}J_{kj,li}-\frac{1}{2}\,n_{ikjl}J_{j,ik} J_{k,jl}-2\, I_{jk} I_{kj}\Big) \,,\\
\label{eq:rel3}K_{i,jk}\overline K_{j,ik}&=I_{ij} J_{k,ij}+2I_{ik} I_{kj}\,, \\
\label{eq:rel4}K_{i,jk}\overline K_{j,lk} &=\sqrt{n_{ijkl}}\Big(I_{kj}J_{lk,ji} + I_{lj} J_{k,ij} \Big)\,,\\
\label{eq:rel5}K_{i,jk} \overline K_{l,ij}&=-\sqrt{n_{ilkj}}\Big(I_{ij} J_{jk,li}+I_{ik} J_{j,il}\Big)\,,\\
\label{eq:rel6}K_{i,jk}\overline K_{j,li}&=\sqrt{n_{ilkj}}\Big(I_{ij} J_{lk,ji}-2 I_{ik}I_{lj}\Big)\,,\\
\label{eq:rel7}K_{i,jk} \overline K_{i,jl}&=-\sqrt{n_{ilkj}} \Big(I_{lj} J_{jk,li}+\frac{1}{2} J_{j,il} J_{lk,ji} \Big)\,.
\end{align}
Another relation is
\begin{equation}\label{eq:rel8}
J_{ji,kl}J_{lk,ij}=4\Big(I_{li}I_{jk}-n_{ikjl}I_{li}I_{jk}+n_{iljk}I_{ji}I_{lk}\Big)+2n_{iljk}\Big(I_{li}J_{jk,li}-I_{jk}J_{li,kj}\Big).
\end{equation}

\end{document}